\documentclass[%
twocolumn,
prx,
]{revtex4-2}

\usepackage{soul}
\usepackage{physics,multirow,siunitx}
\usepackage{graphicx}
\usepackage{dcolumn}
\usepackage{bm}
\usepackage{tikz}
\usepackage{pgfplots}
\usepackage{subcaption}
\usepackage{hyperref}
\usepackage{float}
\usepackage{caption}
\usepackage{ragged2e}
\usepackage{amsmath}
\usepackage{amsfonts}

\begin{document}

\preprint{APS/123-QED}


\title{24 days-stable $\rm{CNOT}$-gate on fluxonium qubits with over 99.9\% fidelity}


\author{Wei-Ju Lin$^{1}$}
\author{Hyunheung Cho$^{1}$}

\author{Yinqi Chen$^{2}$}
\author{Maxim G. Vavilov$^{2}$}
\author{Chen Wang$^{3}$}
\author{Vladimir E. Manucharyan$^{1,4}$}

\affiliation{$^1$Department of Physics, University of Maryland, College Park, MD, USA}
\affiliation{$^2$Department of Physics, University of Wisconsin-Madison, Madison, WI, USA}
\affiliation{$^3$Department of Physics, University of Massachusetts-Amherst, Amherst, MA, USA}
\affiliation{$^4$Institute of Physics, Ecole Polytechnique Federale de Lausanne, Lausanne, Switzerland}

\date{\today}

\begin{abstract}
Fluxonium qubit is a promising elementary building block for quantum information processing due to its long coherence time combined with a strong anharmonicity. In this paper, we realize a 60 ns direct CNOT-gate on two inductively-coupled fluxoniums, which behave almost exactly as a pair of transversely-coupled spin-$1/2$ systems. The CNOT-gate fidelity, estimated using randomized benchmarking, was as high as 99.94\%. Furthermore, the fidelity remains above 99.9\% for 24 days without any recalibration between measurements. Compared with the 99.96\% fidelity of a 60 ns identity gate, our data brings the investigation of the non-decoherence-related errors during logical operations down to $2 \times 10^{-4}$. The present result adds a simple and robust two-qubit gate into the still relatively small family of the “beyond three nines” gates on superconducting qubits.

\end{abstract}

\maketitle

\section{\label{sec:level1}Introduction} 

High-fidelity entangling gate operations is a key requirement for quantum computing \cite{DiVincenzo2000Implementation}. The circuit quantum electrodynamics platform \cite{Blais2021Circuit} relies almost exclusively on transmon qubits \cite{koch2007charge}, where high-fidelity two-qubit gates are implemented using a fast change of qubit parameters \cite{Barends2014Superconducting,barends2019diabatic,Li2019Realisation,negirneac2021high}, all-microwave control \cite{nguyen2024programmable,mitchell2021hardware,wei2022hamiltonian,kandala2021demonstration}, tunable couplers \cite{xu2020high,Stehlik2021Tunable,sung2021realization}, and combinations of the above approaches \cite{Google2020set}. However, all such gate schemes suffer from a relatively weak anharmonicity of transmon circuits, which can lead to leakage errors outside the computational subspace. The leakage erros may be suppressed by replacing transmons with fluxonium qubits \cite{manucharyan2009fluxonium}, which combine long coherence time with a strong anharmonicity \cite{Earnest2018Realization,Lin2018Demonstration,nguyen2019high,Mundada2020Floquet,Zhang2021Universal,somoroff2023millisecond,Sun2023Characterization,Ardati2024Using,mencia2024integer}. Indeed, several high-fidelity two-qubit gates on a pair of fluxonium devices \cite{ficheux2021fast,xiong2022arbitrary,ding2023high,zhang2023tunable,moskalenko2022high,bao2022fluxonium, Ali2024inductive,dogan2023two} all benefit from the strong anharmonicity.

Among the various gate schemes, the cross-resonance (CR) gate \cite{wei2022hamiltonian,Sheldon2016,kandala2021demonstration,Heya2021,Corcoles2013,Hazra2020,Patterson2019} particularly benefits from the high coherence and strong anharmonicity of fluxoniums. First, the CR gate stays dominantly within the fixed-frequency logical subspace and thus may fully harness fluxonium's high coherence. For comparison, the schemes temporarily away from the half-flux "sweet spot" \cite{bao2022fluxonium, moskalenko2022high,Chen2022Fast} or utilizing fluxonium's noncomputational transitions \cite{ficheux2021fast, xiong2022arbitrary, ding2023high} introduce larger incoherent errors. Second, leakage errors limit the recent advances of two-transmon CR gates reaching fidelity up to 99.8\% \cite{wei2022hamiltonian} based on the selective-darkening approach \cite{deGroot2012SD}, while such errors can be largely suppressed in fluxonium systems. Accordingly, applying the selective-darkening variant of the CR gate on fluxoniums would be an attractive scheme chasing high fidelity. 

The selective-darkening variant, also known as direct CNOT \cite{Malekakhlagh2022mitigating,kandala2021demonstration,wei2022hamiltonian,Jurcevic_2021}, was originally implemented for inductively coupled flux qubits \cite{de2010selective}, confining hybridization within computational space and resembling a pair of transversly-coupled spin-1/2. The variant was later applied on the qubits with higher coherence time such as C-shunted flux qubits \cite{Chow2011simple}, capacitively coupled transmons \cite{Malekakhlagh2022mitigating,kandala2021demonstration,wei2022hamiltonian,Jurcevic_2021}, and capacitively coupled fluxoniums \cite{dogan2023two} to achieve high-fidelity gates. However, the circuits directly coupling transmons or fluxoniums via a common capacitor typically induce the repulsion between computational and noncomputational transitions, resulting in quantum crosstalks such as static $ZZ$ interaction and unwanted $ZX$, $XZ$ rates degrading single-qubit operations. This leads to a trade-off between coupling strength (gate speed) and crosstalks, limiting the gate performance \cite{McKay2019Three}. To counter the crosstalks, coupler elements are widely adopted \cite{Stehlik2021Tunable,Houck2011Tunable,Martinis2014Qubit,mckay2016universal,yan2018tunable,Casparis2019Voltage,mundada2019suppression,Ganzhorn2019Gate,Ku2020Suppression,xu2020high,collodo2020implementation,kandala2021demonstration,sung2021realization,ding2023high,zhang2023tunable,moskalenko2022high,bialczak2011fast,Rigetti2018Demonstration,Li2020Tunable,Han2020Error,Kosen2022Building,Warren2023Extensive,song2024realization}. Off-resonantly driving the non-computational transitions offers another strategy canceling static $ZZ$ \cite{Noguchi2020Fast,xiong2022arbitrary} but introduces complexity and additional decoherence. In this work, we inductively couple two fluxoniums acting as two transversely-coupled spin-half systems \cite{couplingscheme} to perform the selective-darkening variant of the CR gate, suppressing quantum crosstalks without additional components.

The simple gate scheme aligns with the goal of pursuing stable gates with high fidelity. There have been a number of demonstrations of two-qubit gates on transmons and fluxoniums with a fidelity approaching 99.9\% \cite{ding2023high,zhang2023tunable,sung2021realization,negirneac2021high,wei2022hamiltonian,Stehlik2021Tunable}. 
However, there are still very few that reliably cross this threshold~\cite{ding2023high, zhang2023tunable}. In this work, we realize a controlled $\pi$ rotation ($CX_{\pi}$) gate (which is equivalent to the CNOT gate after a virtual rotation). The gate fidelity was estimated using conventional randomized benchmarking and its value reaches 99.94\%. More importantly, we show that the value stays above 99.9\% for a period of 24 days. Such a high stability was achieved in part due to a gate scheme without any additional coupler elements or ever activating the non-computational transitions.

This paper is organized as follows. In Section II we introduce our two-fluxonium system using a hardware-independent language of driven coupled two-level systems. The circuit-level analysis and its experimental verification is presented separately in Ref.~\cite{couplingscheme}. In Section III we describe the benchmarking of the selective darkening/cross-resonance gate. The details of the experiment, including the tune-up procedures are explained in the Appendices. The paper is concluded with a discussion of errors (Section IV) and a short summary (Section V).

\section{\label{sec:level1}System description\protect\\} 

\begin{figure*} 
    \centering
    \begin{tikzpicture}
        \node[inner sep=0] (image1) at (-12,3.8) {\includegraphics[width=0.45\textwidth]{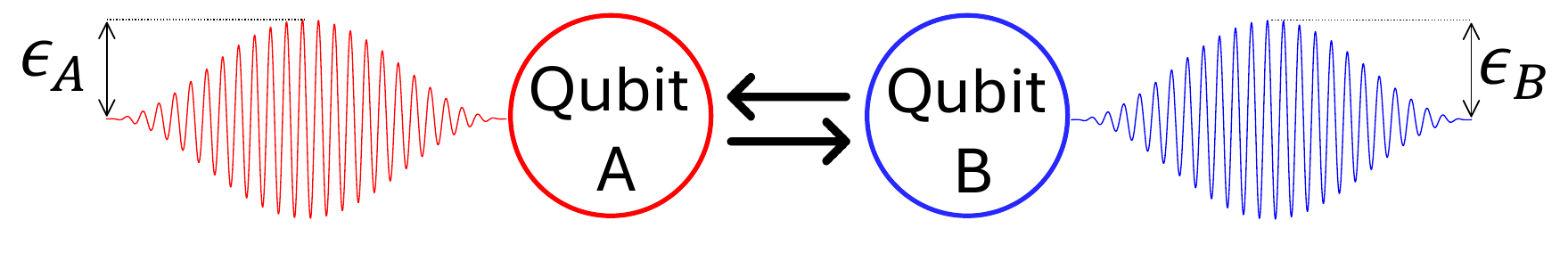}};
        \node at (-15.5,5) {(a)}; 

        \node[inner sep=0] (image2) at (-3,2.3) {\includegraphics[width=0.55\textwidth]{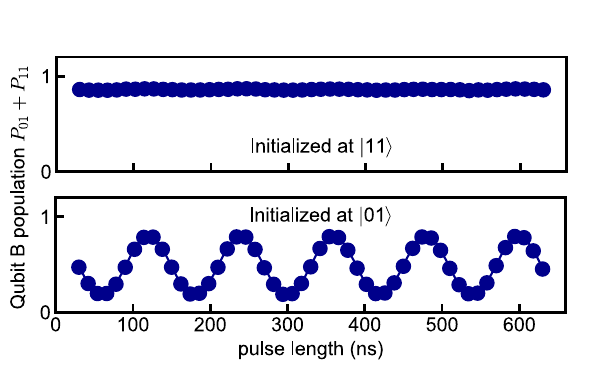}};
        \node at (-15.5,3) {(b)}; 
        \begin{scope}[xshift=-12.5cm, yshift=-0.4cm]
          \draw[line width=0.6mm] (1,0)  -- (2.5,0) node[right, font=\fontfamily{phv}\selectfont] {$\mathsf{|00\rangle}$};
          \draw[dotted] (-1, 0) -- (1, 0);
          \filldraw[black] (-1,0) circle (2pt) node[left, font=\fontfamily{phv}\selectfont] {0 GHz};
          \draw [line width=0.6mm](0,1) node[left, font=\fontfamily{phv}\selectfont] {$\mathsf{|10\rangle}$}-- (1.5,1) ;
          \draw[dotted] (-1, 1) -- (0, 1);
          \filldraw[black] (-1,1) circle (2pt) node[left, font=\fontfamily{phv}\selectfont] {0.15 GHz};
               
          \draw [line width=0.6mm](1,2.5) -- (2.5,2.5) node[right, font=\fontfamily{phv}\selectfont] {$\mathsf{|11\rangle}$};
          \draw[dotted] (-1, 2.5) -- (1, 2.5);
          \filldraw[black] (-1,2.5) circle (2pt) node[left, font=\fontfamily{phv}\selectfont] {0.38 GHz};
          \draw [line width=0.6mm](1.8,1.5) -- (3.3,1.5) node[right, font=\fontfamily{phv}\selectfont] {$\mathsf{|01\rangle}$};
          \filldraw[black] (-1,1.5) circle (2pt) node[left, font=\fontfamily{phv}\selectfont] {0.23 GHz};
          \draw[dotted] (-1, 1.5) -- (2, 1.5);
          \draw[thick,blue, dashed] (2.5,0.75) arc[start angle=0,end angle=185,x radius=0.2, y radius=0.75] ;
          \draw[->,>=stealth,thick,blue] (2.3,0) arc[start angle=-90,end angle=70,x radius=0.2, y radius=0.75] node[right, pos=0.55, font=\fontfamily{phv}\selectfont] {$\mathsf{CX_{\pi}}$};
          \draw[blue, thick, dashed] (2.1,0.75) arc[start angle=-180,end angle=10,x radius=0.2, y radius=0.75 ];
          \draw[thick, blue, dotted] (1.5,0.57+1.18) arc[start angle=0,end angle=180,x radius=0.2, y radius=0.75]  node[left, pos=0.9, align=left, font=\fontfamily{phv}\selectfont] {Selectively\\Darkened};
          \draw[thick, blue, dotted] (1.1,0.57+1.18) arc[start angle=-180,end angle=0,x radius=0.2, y radius=0.75];
          \draw[->, thick] (-1,0) -- (-1,3) node[above] {$f$};
    \end{scope}
        \node at (-8,5) {(c)}; 
    \end{tikzpicture}
    \caption{\label{fig:LevelsDriveCRabi} \justifying (a) Illustration of two local charge drives for various gate operations. (b) Energy levels of our two-qubit system. The blue lines illustrate the scheme of the two-qubit $CX_\pi$ gate based on selective darkening. (c) Conditional Rabi oscillation of qubit B initialized at $|11\rangle$ and $|01\rangle$, comparing the dynamics of the dark ($|10\rangle$-$|11\rangle$) and bright ($|00\rangle$-$|01\rangle$) transitions under selective darkening CR drive.}

\end{figure*}
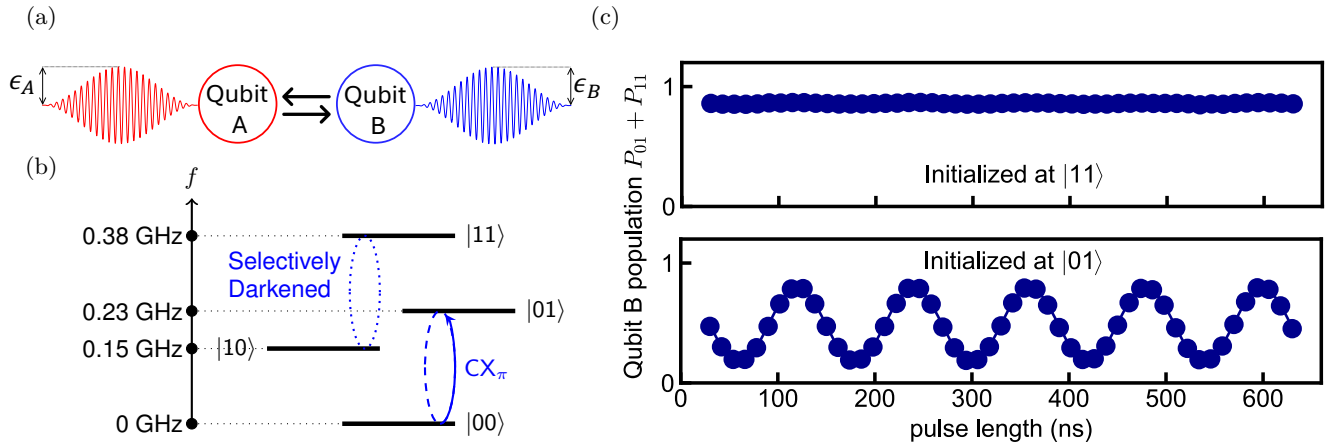

Our device consists of two fixed-frequency fluxoniums that exhibit the behavior of a transversely coupled spin-1/2 system. The system's lowest energy states serve as the computational basis $|00\rangle$, $|10\rangle$, $|01\rangle$, and $|11\rangle$, revealing the symmetric nature of two coupled spin-1/2 as illustrated in the next section. We apply a microwave drive to manipulate our qubits as shown in Fig. \ref{fig:LevelsDriveCRabi}(a). Figure~\ref{fig:LevelsDriveCRabi}(b) shows the qubit frequencies and level structures of the system.

\subsection{\label{sec:level2}Fomulation of Hamiltonian}

We consider two coupled qubits under a monochromatic charge drive and model this system by the Hamiltonian restricted to the computational subspace. This Hamiltonian is consistent with the inductively coupled fluxonium system discussed in \cite{couplingscheme}. The low-energy part of the Hamiltonian in the computational subspace can be written as
\begin{equation}
    \hat{H}_{\mathrm{sys}} / h = 
    \frac{f_{01}^A}{2} \hat{\sigma}_z^A+\frac{f_{01}^B}{2} \hat{\sigma}_z^B+\frac{\xi_{Z Z}^{static}}{4} \hat{\sigma}_z^A \hat{\sigma}_z^B + \hat{H}_{\text{drive}}/h,
    \label{eq:totalHamiltonian}
\end{equation}
where $h$ is the Planck constant, $f_{01}^j$ is the transition frequency of qubit $j$, $\hat{\sigma}_z^j$ represents the $Z$ Pauli matrix for qubit $j$. $\xi_{Z Z}^{static}$ is the static ZZ phase accumulation rate. The microwave drive in the computational subspace reduces to
\begin{equation}\label{eq:DriveEffHamiltonian}
    \begin{aligned}
        \hat{H}_{\text {drive}}/h= & \beta(t)
        \left(\xi_{A}^{+} \hat{\sigma}_x^A+\xi_{A}^{-} \hat{\sigma}_x^A\hat{\sigma}_z^B
        +\xi_{B}^{+} \hat{\sigma}_x^B+\xi_{B}^{-} \hat{\sigma}_z^A\hat{\sigma}_x^B\right),\\
        \beta(t) = & \Bar{\beta}(t) \cos \left(2 \pi f_d t+\phi\right). 
    \end{aligned}
\end{equation}
Here, $\beta(t)$ describes the time dependence of the drive determined by the drive frequency $f_d$, the phase $\phi$, and the slow-varying drive-envelop function $\Bar{\beta}(t)$. $\hat{\sigma}_x^j$ stands for the $X$ Pauli matrix for qubit $j$. $\xi_{A}^{+}$, $\xi_{A}^{-}$, $\xi_{B}^{+}$, and $\xi_{B}^{-}$ are the rates of effective $XI$, $XZ$, $IX$, and $\epsilon_{A}$ and $\epsilon_{B}$ applied to fluxonium A and B as   
\begin{equation}
\label{eq:xi_A_pm}
    \left[\begin{array}{l}
         \xi_{A}^{\pm}  \\
         \xi_{B}^{\pm}
    \end{array}\right]
    =
    \left[\begin{array}{cc}
        M_{AA}^\pm & M_{AB}^\pm  \\
        M_{BA}^\pm &  M_{BB}^\pm 
    \end{array}\right]
    \left[\begin{array}{l}
         \epsilon_{A}  \\
         \epsilon_{B}
    \end{array}\right]\,.
\end{equation}
Here, $\epsilon_{A}$ and $\epsilon_{B}$ are real coefficients proportional to the in-phase drive amplitudes we apply to qubits A and B, respectively. These local amplitudes take into account the combined effects of classical crosstalks and the two port drive amplitudes, which are discussed in Appendix \ref{Classical crosstalks}. Coefficients $M_{kl}^\pm$ connect the interaction rates $\xi_{k}^{\pm}$ and the drive amplitudes $\epsilon_{l}$ with $k,l=\{A,B\}$. The values of $M_{kl}^\pm$ computed for our system are provided in Table~\ref{tab:MatrixElements}, and the origin of these coefficients is discussed in Appendix~\ref{app:Happ}.

\begin{table}
  \caption{\label{tab:MatrixElements} \justifying Coefficients to connect $\xi_{k}^{\pm}$ and $\epsilon_l$ with $k,l=A,B$.}
  \begin{tabular}{l|cc|| l|cc}
  \hline \hline \multirow{1}{*}{$M_{kl}^+$} & $l=A$ & $l=B$ &  $M_{kl}^-$ & $l=A$ & $l=B$ \\
  \hline
  $k=A$& 0.093736 & 0.000026  &
  $k=A$& -0.000006 & -0.036768 \\  
  $k=B$& 0.000025 & 0.127758 &
  $k=B$& 0.041791 & 0.000012\\
  \hline \hline
  \end{tabular}
\end{table}

Within the CR gate scenario, we can apply any linear combination of the two local drives, $\epsilon_A$ and $\epsilon_B$, at the frequency $f_{01}^A$ or $f_{01}^B$, thus inducing resonant transitions in qubit A or B, respectively.
To have a better understanding of our system, we can see $M_{AA}^-$, $M_{BB}^-$, $M_{AB}^+$, and $M_{BA}^+$ are all close to 0, meaning that we have four simple control knobs: (i) the local drive $\epsilon_A$ at the frequency $f_{01}^A$, (ii) the local drive $\epsilon_B$ at the frequency $f_{01}^A$, (iii) the local drive $\epsilon_B$ at the frequency $f_{01}^B$, and (iv) the local drive $\epsilon_A$ at the frequency $f_{01}^B$. These control knobs provide tuning the $XI$, $XZ$, $IX$, and $ZX$ interaction rates individually, facilitating a simple scheme for gate operations. Note that the CR drive $\epsilon_B$ ($\epsilon_A$) at the frequency $f_{01}^A$ ($f_{01}^B$) corresponds to $M_{AB}^-$ ($M_{BA}^-$), showing magnitude comparable to $M_{AA}^+$ ($M_{BB}^+$). This indicates the strong hybridization of our computational space, providing a large $XZ$ $(ZX)$ rate under CR drive that is comparable to the $XI$ $(IX)$ rate of single-qubit gates. Even though the computational states are strongly hybridized, a single local drive $\epsilon_A$ ($\epsilon_B$) at the frequency $f_{01}^A$ ($f_{01}^B$) synchronizes single-qubit gates without $XZ$ $(ZX)$ interaction. The elements with magnitude well below $10^{-4}$ in Table~\ref{tab:MatrixElements} make our device suitable for single and two-qubit gates and is consistent with the features of transversely coupled spin-1/2 systems.

\subsection{\label{sec:level2}System characterization}

Our device with two control lines that are charge coupled to each qubit is embedded into a 3D copper cavity resonator to perform single-shot joint readout. We fix the external magnetic field to bias qubits at the flux-insensitive sweet spot. The measured coherence times and single-qubit Clifford-gate fidelities, $F_{I}$ and $F_{s}$, extracted from individual and simultaneous randomized benchmarking (RB) for each qubit are presented in Table~\ref{tab:table1}, which summarizes the key parameters of our fluxonium qubits. Note that the small static $ZZ$ values under such a strong qubit-qubit coupling agrees well with the purely two level model. The calibration process and benchmarking of single-qubit gates are shown in Appendix \ref{sec:1QBcal}. The characterization and extraction of fluxonium parameters is demonstrated in \cite{couplingscheme}, which also discusses the circuit nature causing the suppressed $ZZ$ rate. The details of experimental procedures, such as initialization and readout, are provided in Appendix \ref{sec:exsetup}.

\begin{table}
  \caption{\label{tab:table1}\justifying Fluxonium qubit parameters and performance. $\alpha$ represents anharmonicity. $F_{I}$ and $F_{s}$ are the fidelities extracted from individual and simultaneous randomized benchmarking, respectively. }
  
  \begin{tabular}{lcc}
  \hline \hline & \text{Qubit A} & \text{Qubit B} \\
  \hline
  $f_{01} (\mathrm{GHz})$ & 0.15 & 0.23 \\
  $\:\: \alpha \:(\mathrm{GHz})$ & 4.51 & 4.54 \\
  $\xi_{Z Z}^{static} (\mathrm{MHz})$ & \multicolumn{2}{c}{0.002} \\
  $\:\:T_1 (\mathrm{\mu s})$ & 260 & 160 \\
  $\:\:T_2^* (\mathrm{\mu s})$ & 100 & 110 \\
  $\:\:T_2^E (\mathrm{\mu s})$ & 200 & 150 \\
  $\:\:F_{I} (\%)$ & 99.976 & 99.96 \\
  $\:\:F_{S} (\%)$ & 99.93 & 99.913 \\
  \hline \hline
  \end{tabular}
\end{table}

\section{two-qubit GATE concept, TUNING, AND RESULTS}

\begin{figure*}
       
    \includegraphics[width=\textwidth]{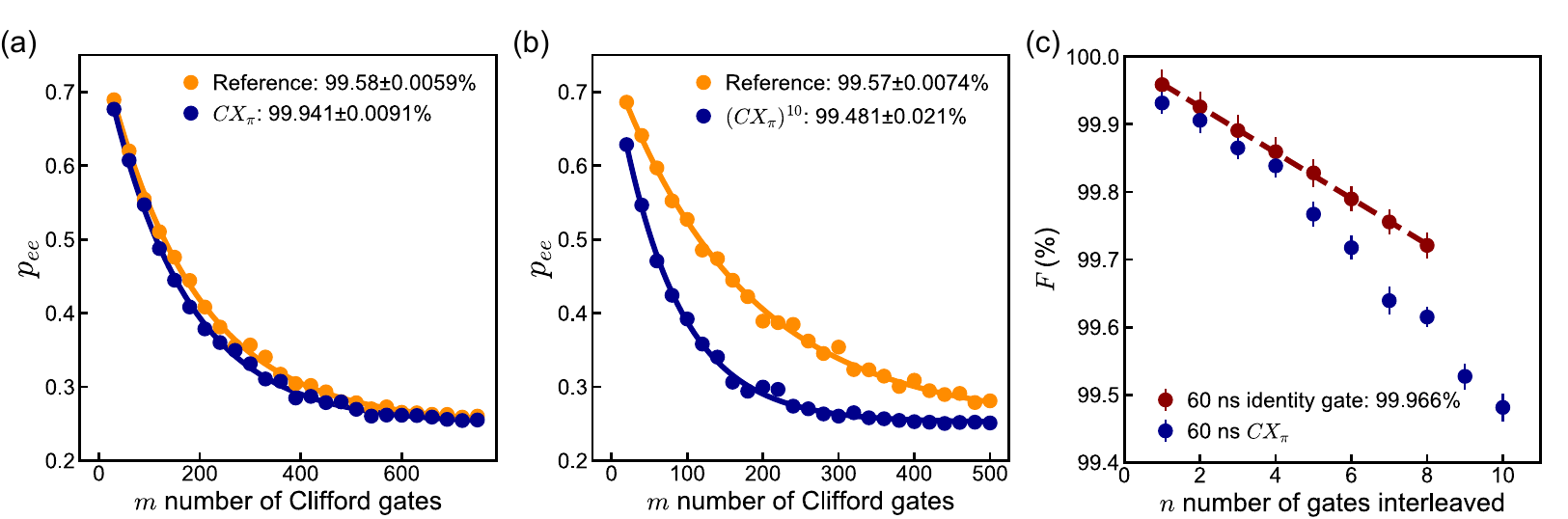}
    \caption{\label{fig:RB_traces} \justifying IRB results. (a), (b) IRB of the 60 ns $CX_{\pi}$ with $n=1,10$, respectively. We construct the two-qubit Clifford group with, on average, 4.5 physical single-qubit gates (incorporating virtual Z gates) and 1.5 $CX_\pi$ gates per Clifford gate. The reference error per two-qubit Clifford extracted from the standard RB is consistent with the average number of gates and the corresponding gate infidelity. (c) Fidelities of $CX_\pi$ and idling identity gates with different $n$. Here, we show the fidelity number extracted from the linear fit for the 60 ns identity gate.}
\end{figure*}

\subsection{Gate concept}

\subsubsection{$ZX_{-\pi/2}$ gate}
In this section, we specify the essential conditions of the two CR gates, direct $CX_\pi$ based on selective darkening and $ZX_{-\pi/2}$.
For the description later in this section, we label the control qubit as A and the target qubit as B.
In the $ZX_{-\pi/2}$ scheme, the two target qubit transitions rotate by the same angle but with opposite directions conditional on the control qubit state.  The latter is achieved  if $\langle 00 | \hat{H}_{\text {drive }} | 01 \rangle=-\langle 10 | \hat{H}_{\text {drive }} | 11 \rangle$. For $CX_\pi$ gate based on selective darkening, we need to darken one of the target qubit transitions.  In this paper, we choose $\langle 10 | \hat{H}_{\text {drive }} | 11 \rangle=0$,  as illustrated in Fig. \ref{fig:LevelsDriveCRabi}(b).

We provide a perspective on the gate operations using expression~\ref{eq:DriveEffHamiltonian}. To perform $ZX_{-\pi/2}$ gate, we drive the system resonantly at the transition frequency $f_{01}^B$ of qubit B. We choose the drive amplitudes $\epsilon_{A,B}$ so that 
$\xi_{B}^{+}$ is strongly reduced. Thanks to the tiny value of the coefficient $M_{BA}^+$, see Table~\ref{tab:MatrixElements}, this requirement is satisfied if $\epsilon_B=0$ and $\epsilon_A \neq  0$. The conditional rotation of the target qubit is then achieved due to the resonant drive of qubit B transition with term $\xi_B^- = M^-_{BA}\epsilon_A$.  Note that the terms driving the control qubit, $\xi^\pm_A$, do not vanish in this case and are responsible for the undesirable transitions of the control qubit.  However, the contribution to the gate error for these transitions remain below $10^{-4}$ for gate times exceeding 50 ns. 

\subsubsection{$CX_\pi$ gate}
The $CX_\pi$ gate with $\langle 10 | \hat{H}_{\text {drive }} | 11 \rangle=0$ requires $\xi_{B}^{-}=\xi_{B}^{+}$, which is satisfied by choosing $\epsilon_A/\epsilon_B \approx M^+_{BB}/M^-_{BA}$.  This condition for our system is satisfied when the amplitudes $\epsilon_{A,B}$ are comparable to each other and do not require an irradiation of one of the qubits with extremely strong drive to have the gate  $t_{\rm gate}$ determined by the drive amplitude according to the following relation $\xi_{B}^{-}=\xi_{B}^{+} \propto 1/t_{\mathrm{gate}}$.  The off-resonant drive for the control qubit is defined by $\xi_A^\pm$ coefficients and cause transitions with relatively low probabilities for the gates longer than 50 ns, see Appendix~\ref{app:budget}.  

\subsubsection{Single-qubit gates}

Next, we argue that our device is also suitable for single-qubit gates that require $\xi_{A,B}^-=0$ and $\xi_{A,B}^+\neq 0$.  The values of coefficients $M^\pm_{kl}$ in Table~\ref{tab:MatrixElements} shows that a direct resonant drive of qubit A ($\epsilon_A\neq 0$ and $\epsilon_B = 0$) or qubit B ($\epsilon_A = 0$ and $\epsilon_B \neq 0$) will generate a rotation of the quantum state in the corresponding qubit computational subspace independent of the other qubit state. 

We mention that our system is not restricted to selecting a particular fluxonium pair as the control-target qubits. The description above also applies to the gate operations driving at $f_{01}^A$. For example, our $ZX_{-\pi/2}$ experiments discussed in Appendix \ref{Sec:ZX-90} utilize qubit A(B) as the target(control) qubit. In this case, the essential condition for the $ZX_{-\pi/2}$ gate is $\xi_{A}^{+}=0$.

\subsection{Gate calibration}

Our drive frequency $f_d$ is on-resonant with the frequency of the bright transition $|00\rangle-|01\rangle$, which is extracted from the Ramsey measurement. To implement high-fidelity gates, we next iterate through tune-up sequences that are sensitive to specific errors and calibrate against the target parameters with a large number of pulses to amplify the error for precise tuning. The seven fine-tuned target parameters of our CR pulse are listed below along with the corresponding goals for a $CX_\pi$ gate as an example: the relative amplitude and phase of the two port drives to selectively darken the $|10\rangle$-$|11\rangle$ transition, the DRAG coefficient \cite{Gambetta2011DRAG,Motzoi2009DRAG} with fixed drive frequency to achieve a big circle trajectory on the Bloch sphere, the overall drive amplitude to ensure a $\pi$ rotation angle, the common pulse phase to align the rotation axis to $X$ rotation, and the single-qubit phase accumulation of A and B due to the CR pulse to determine the phase values of virtual Z \cite{McKay2017VZ} compensation. The whole tuning process takes roughly three hours. Appendix \ref{sec:CXcal} describes the time cost of each step and demonstrates the detailed procedure of $CX_\pi$ gate tuning. Similar tune-up procedures for direct CNOT gates have been reported in \cite{dogan2023two},\cite{wei2022hamiltonian}. The tuning of $ZX_{-\pi/2}$ gate is provided in Appendix \ref{Sec:ZX-90}.

\begin{figure*}
    \includegraphics[width=\textwidth]{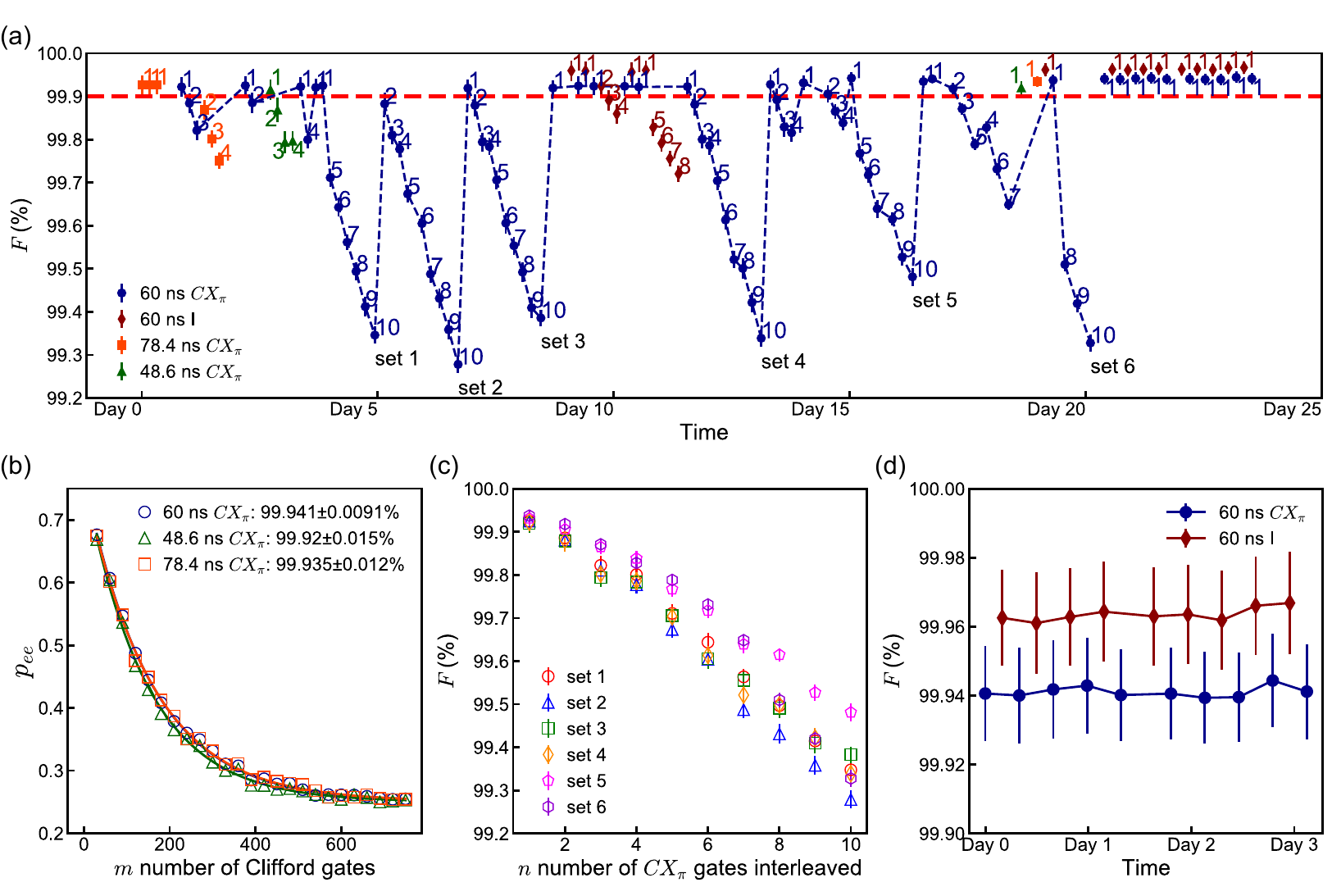}
    \caption{\label{fig:RB_stability} \justifying IRB measurements spanning over 3 weeks without any recalibration, demonstrating the stability of $CX_\pi$ gates. (a) All measured IRB fidelities with an auxiliary line indicating 99.9\%. The numbers next to the data points represent the value of $n$. Our fridge warmed up due to a power outage on Day 25, which ended the measurement. (b) IRB traces interleaved with different gate times of $CX_\pi$ along with the fidelity numbers. (c) Comparison of all 6 sets of fidelities versus $n$. The fifth set reaches the highest fidelities, coming from a gradual change near Day 15. Starting from the eighth data point of set 6, fidelities drop back to the values of the previous sets, happening right after the reboot of arbitrary waveform generator (AWG). Thus the fluctuation of fidelities may relate to the instability of AWG. (d) Interleaving $CX_\pi$ IRB with idling identity gate IRB. The $CX_\pi$ and identity gates have average fidelity above 99.94\% and 99.96\%, respectively. This difference of $2 \times 10^{-4}$ sets the upper bound for coherent errors. }
\end{figure*}

\subsection{Gate characterization}

We perform interleaved randomized benchmarking (IRB) \cite{Corcoles2013} to characterize our CR gates with an offset cosine pulse shape without flattop. We focus on the characterization of $CX_\pi$ in this section, while the IRB results of $ZX_{-\pi/2}$ are presented in Appendix \ref{Sec:ZX-90}. Figure\ref{fig:RB_traces}(a) shows the results of 100 randomized gate sequences for a 60 ns $CX_\pi$ gate with the extracted infidelity of $6 \times 10^{-4}$. To better characterize the stability of gates, we increase the number of $CX_\pi$ gates interleaved $n$ up to 10 to amplify the error. Fig. \ref{fig:RB_traces}(b) shows the IRB traces of the $n=10$ case for example. The corresponding infidelity is much larger than the error bar, enabling us to monitor the gate performance over time without ambiguity, as shown in the next paragraph.  
We plot the fidelities versus $n$ for the 60 ns idling identity and $CX_\pi$ gates in Fig. \ref{fig:RB_traces}(c) and extract the fidelity of the identity gate through linear fit.

In Fig. \ref{fig:RB_stability}(a), we showcase all the IRB measurements spanning over three weeks without any recalibration interleaved to demonstrate the stability of our device. We simply let a measurement computer repeatedly recall the same configuration files to perform the same IRB with all parameters being the same. In this work, we calibrate three different gate times of $CX_\pi$ as shown in Fig. \ref{fig:RB_stability}(b), all reaching fidelity higher than 99.9\%. In Fig. \ref{fig:RB_stability}(c), 6 sets of IRB measurements for 60 ns $CX_\pi$ versus $n$ are presented together, giving the information of repeatability within 20 days. By interleaving $CX_\pi$ gate IRB with identity gate IRB as shown in Fig. \ref{fig:RB_stability}(d), we demonstrate that our $CX_\pi$ is mostly limited by incoherent errors. The two gate fidelities are measured alternately on the time scale of days, providing statistics that minimize the effect of coherence time fluctuation. The difference between the two fidelities points out that the additional errors are on the order of $2 \times 10^{-4}$, setting a upper bound for coherent errors.

\subsection{Error budget} \label{Sec:error budget}

\begin{figure}
    \includegraphics[width=0.4\textwidth]{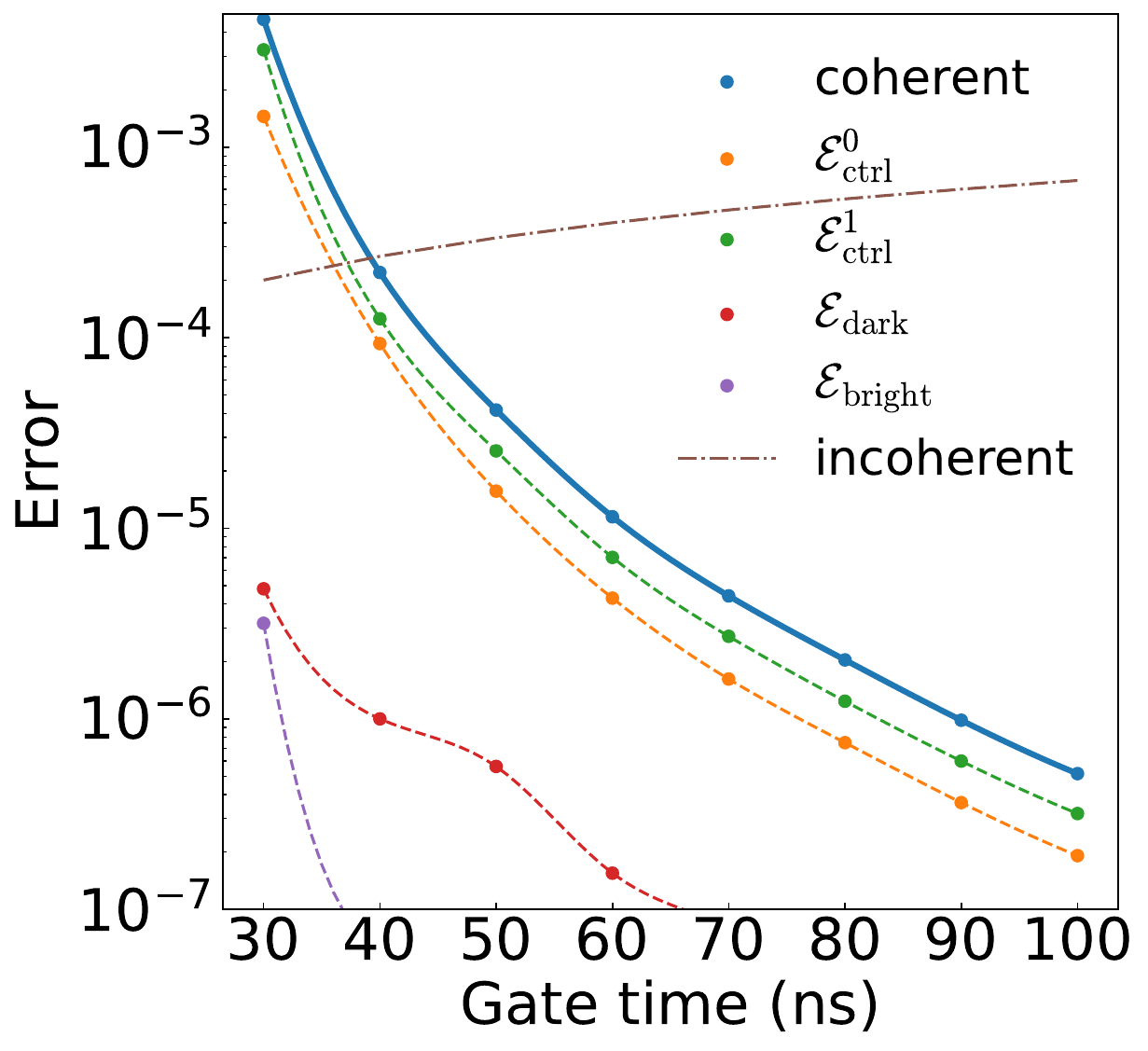}
    \caption{\justifying Error budget of the $CX_\pi$ gate for different gate times. The solid line is the coherent error of the optimized gate. The dashed lines are the decomposition of the coherent error into different error channels. The errors, $\mathcal{E}_{\mathrm{ctrl}}^0$ ($\mathcal{E}_{\mathrm{ctrl}}^1$), $\mathcal{E}_{\mathrm{dark}}$, and $\mathcal{E}_{\mathrm{bright}}$, represent the unwanted excitation of the control qubit transition when target qubit is in state $\ket{0}$ ($\ket{1}$), the dark transition, and the bright transition, respectively. These values are determined by the matrix elements of the simulated evolution operator with further details provided in Appendix \ref{app:budget}. The dotted-dashed line is the analytical estimation of the incoherent error using Eq. \ref{eq:IncoherentError}.}
    \label{fig:cnot_sim}
\end{figure} 

To investigate the infidelity of the $CX_\pi$ gates on our device, we run simulations evaluating the sources of coherent errors for different gate times as shown in Fig. \ref{fig:cnot_sim}. While it is preferred to have as short a gate time as possible to reduce the effect of decoherence. As shorter gate time requires a stronger drive, there is a lower bound on the gate time due to control errors. In our case, the dominating error originates from control qubit flips that happen even though the drive frequency is detuned from the control qubit frequency. Such a control qubit flip error dominates the coherent error of our gate, reaching $10^{-5}$ with a gate time of 60 ns. The error can further exceed $10^{-3}$ by shortening the gate below 30 ns. The other coherent errors, including the bright and dark transition errors and the leakage error, have a negligible contribution to the total gate error. We present the details of the simulation for error budget in Appendix~\ref{app:budget}.

\section{Discussion}

Recent works have adopted inductively coupled fluxoniums for two qubit gates \cite{Ali2024inductive,zhang2023tunable}, utilizing flux drives for gate operations. The first report \cite{Ali2024inductive} performs rf dynamical decoupling drive to counter the dephasing due to flux noise and improve coherence times with an order of magnitude, achieving 20 ns fast gates but still with a coherence time drop when the flux drive is applied. The fluxonium qubits of the latter work \cite{zhang2023tunable} are coupled through a tunable coupler, consisting of a static harmonic mode and a flux tunable fluxonium-like mode. Modulation of coupler flux provides $XX$ interaction between qubits for two-qubit gates, which can be turned off by making the contributions from the two modes equal magnitude but opposite signs, enabling on-off control on qubit-qubit interactions. In comparison, we perform charge drive without tuning magnetic flux, exploring different gate controls for inductively coupled fluxonium qubits. Despite that fixing qubits at the sweet spot minimizes the sensitivity to flux noise, charge controls drive high transitions stronger compared with flux modulation, giving higher possibility of leakage and requiring more stringent criteria for frequency allocation while designing circuits.

Charge driven gates on fixed frequency fluxonium qubits with fidelity above 99.9\% and gate times similar to ours are reported in \cite{ding2023high}. The microwave activated $CZ$ gate comes from a full oscillation of higher transitions. The transmon coupler enables individual controls with minimized crosstalks. The gate tuning using reinforcement learning mitigates coherent errors larger than $5 \times 10^{-4}$, showing significant suppression of this error. To compare, our gates stay within computational subspace based on tune-ups without any black-box process such as Nelder-Mead or machine learning algorithms, suppressing coherent errors down to below $2 \times 10^{-4}$. This process provides transparency with short time cost but only deals with limited numbers of error types. Here, we stress that the tune-up sequences are all designed against the coherent errors within computational space. That is, although we do not aim at the parameters that can mitigate leakage errors, it's still lower than $2 \times 10^{-4}$, thanks to the large anharmonicity. In this sense, our system is favorable for experimental quantum error correction compared with the limiting leakage error of the CR gates on transmons.

Both the qubit-qubit coupling scheme discussed in \cite{couplingscheme} and the gate scheme presented in this letter are all confined within the lowest four states, designed to fulfill the full potential of fluxonium's high coherence computational space without quantum crosstalks. The elements with magnitude well below $10^{-4}$ in Table~\ref{tab:MatrixElements} of our strongly coupled fluxonium qubit device illustrate the characteristics of a transversely coupled spin-1/2 system, providing simple control protocols and separating this work from the previous CR gate on fluxonium qubits reported in \cite{dogan2023two}.  

The freedom to choose a $CX_\pi$ or a $ZX_{-\pi/2}$ gate manifests the versatility of the CR gate family, while the difference between these two options for our system can be briefly discussed below: The $ZX_{-\pi/2}$ is subject to additional coherent errors such as dynamical $ZZ$ errors in comparison with the $CX_\pi$, for which we apply single-qubit virtual $Z$ gates to compensate $ZZ$ dynamics as illustrated in Appendix \ref{sec:CXcal}. On the other hand, the $ZX_{-\pi/2}$ of our system requires only a single local drive on the control qubit, reducing the technical complexity of gate operations.

We next want to comment on the difference between the simulated $10^{-5}$ coherent error and the measured upper bound of $2 \times 10^{-4}$ for coherent errors. According to our analysis, the drive envelope delay due to the cable length difference between the two port drives is probably the main reason. Fig. \ref{fig:envelope_delay} in Appendix \ref{app:budget} provides the corresponding details, showing the importance of synchronizing drive envelopes especially for coherent errors below $10^{-4}$. 

We next discuss the discrepancy of individual and simultaneous single-qubit RB measurements. The resonant direct drive terms in Hamiltonian \ref{eq:DriveEffHamiltonian} defined by coefficients $M^{+}_{AA}$ and $M^{+}_{BB}$ provide the leading contribution to single-qubit gates.
The contribution from the non-resonant cross-drives is considered to be small. In fact, the qubit-qubit detuning $\delta=80$ MHz is relatively close to the Rabi frequency $\propto 1/t_{\textrm gate}$, and the cross resonance coefficients $M^{-}_{AB},M^{-}_{BA}$ in Table~\ref{tab:MatrixElements} are only a few times smaller than the direct coefficients $M^{+}_{AA}$ and $M^{+}_{BB}$. This indicates the contribution of cross-resonant terms in Hamiltonian \ref{eq:DriveEffHamiltonian} has to be taken into account.
The corresponding Stark shifts further degrade the performance of simultaneous gates. To investigate this effect, we simulate the coherent errors of the individual X gates on each qubit and the simultaneous X gate pair on both qubits in Appendix \ref{app:budget}. The degradation of single-qubit gates due to stronger hybridization manifests another trade-off relation between single-qubit and two-qubit gates while tuning the qubit-qubit coupling strength. Future devices with larger qubit-qubit detuning can mitigate this effect for the same Rabi frequency and, thus, the same gate time. Since the hybridization ratio is proportional to $J_L/|f_{01}^{A} - f_{01}^{B}|$, the smaller ratio of cross resonance coefficients to direct coefficients ($M^{-}_{AB}/M^{+}_{AA}, M^{-}_{BA}/M^{+}_{BB}$) due to larger detuning can be tuned back to the same ratio with stronger qubit-qubit coupling constant. In this case, we would get the same two-qubit gate speed and better performance of simultaneous single-qubit gates. The larger detuning will also benefit the two-qubit $CX_\pi$ gate by reducing the control qubit errors, allowing shorter gate times. Thus, the larger qubit-qubit detuning enables the possibility of better performance of single and two qubit gates and is possible considering the unused frequency window with a range of a few hundred MHz. Designing the level structures and gate schemes as simple as possible, the proposed high-fidelity gates with maximized stability are suitable for practical applications in the future.

\section{Summary}

We demonstrate a 60 ns direct $CX_{\pi}$ gate with fidelity reaching 99.94\%. The fidelity maintains stability for over 3 weeks without recalibration. The analogy of transversely coupled spin-1/2 systems relaxes the trade-off condition between coupling strength and crosstalks, which is limiting most of the present schemes utilizing direct coupling. Another trade-off relation between two-qubit and single-qubit gates while tuning the hybridization strength can also be relaxed with larger qubit-qubit detuning for future devices. The weak quantum crosstalks of our stronly hybridized qubit transitions, the low leakage error without corresponding tune-ups, along with the exceptional stability and simplicity of our gates experimentally illustrate a new pathway for scaling up and quantum error correction, following the previous theory proposal \cite{nguyen2022blueprint}. However, 
future devices coupling more qubits still need to address classical crosstalks, which usually come after the typically short distances between the qubits sharing common junctions. Long range coupling using loop-type mutual inductance without introducing classical crosstalks and extra degrees of freedom in the circuit \cite{couplingscheme} is desired.

\begin{acknowledgments}

We acknowledge Lincoln Labs and IARPA for providing Josephson Traveling Wave Parametric Amplifiers to enhance the readout. This research was supported by the ARO HiPS (contract No. W911-NF18-1-0146) and GASP (contract No. W911-NF23-10093) programs.

\end{acknowledgments}

\appendix

\section{Experimental setups and methods} \label{sec:exsetup}

\begin{figure}[b]
\includegraphics[scale=0.90]{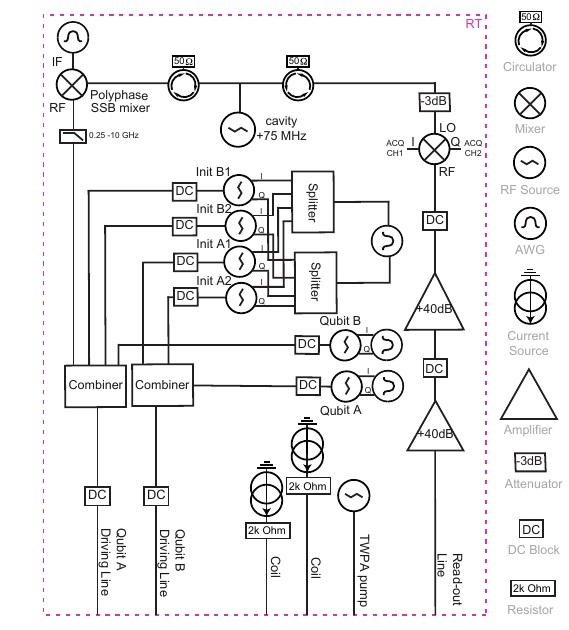}
\caption{\label{fig:exsetup} Schematics of experimental setup.}
\end{figure}

The measurement setups are similar to those previously reported in \cite{xiong2022arbitrary}, and we show the difference especially the room temperature setup in Fig. \ref{fig:exsetup}. We send external driving into the cavity through two input ports and measure the transmission signal with the third port. The device image can be found in \cite{couplingscheme}.

\begin{figure*}
    \centering
    \begin{tikzpicture}
        \node[inner sep=0] (image1) at (-1,0) {\includegraphics[width=0.45\textwidth]{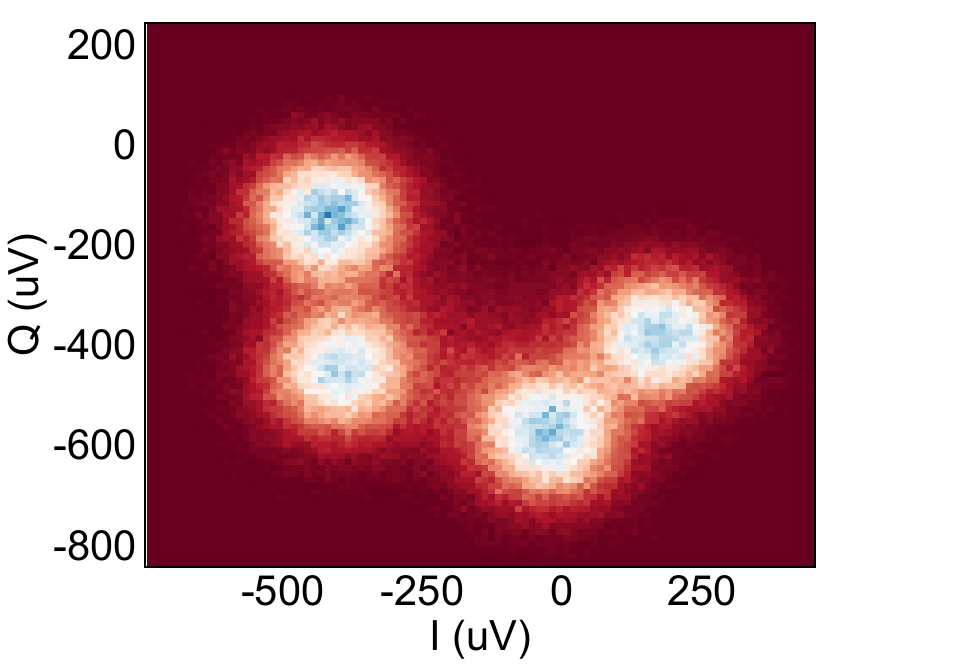}};
        \node at (-4.8,3) {(a)}; 

        \node[inner sep=0] (image2) at (5.5,0) {\includegraphics[width=0.4\textwidth, height=0.31\textwidth]{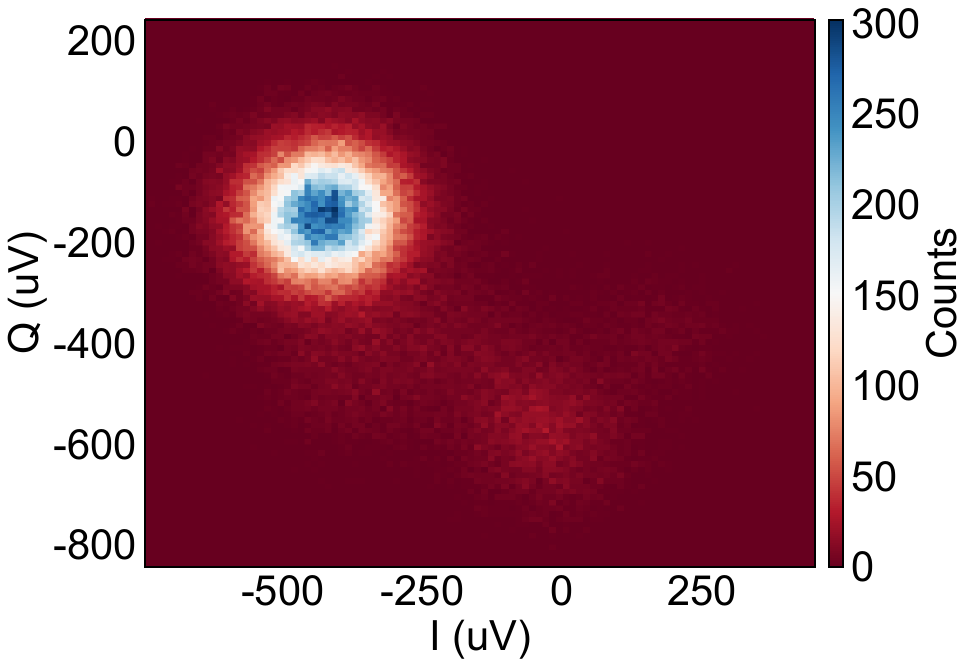}};
        \node at (2.1,3) {(b)}; 

        \node[inner sep=0] (image3) at (11,0) {
            \begin{tikzpicture}[scale=1, xscale=0.65]
                \draw[line width=0.6mm] (0,0.5) node[left] {$|g0\rangle$} -- (1.5,0.5);
                \draw[line width=0.6mm] (0,1) node[left] {$|e0\rangle$} -- (1.5,1);
                \draw[line width=0.6mm] (0,3) node[left] {$|f0\rangle$} -- (1.5,3);
                \draw[line width=0.6mm] (0,3.5) node[left] {$|h0\rangle$} -- (1.5,3.5);
                \draw[->,>=stealth,thick,orange] (0.75,0.5) -- (0.75,3.5);

                \draw[line width=0.6mm] (2,5.5) -- (3.5,5.5) node[right] {$|e1\rangle$};
                \draw[->,>=stealth,thick,red] (0.75,3.5) -- (3,5.5);
                \draw[<-,>=stealth,thick,blue] (0.75,1) -- (3,5.5);

                \begin{scope}[xshift=2.3cm, yshift=3.3cm]
                    \draw[thick, blue] (0,0) sin (0.0625,0.5) cos (0.125,0) sin (0.1875,-0.5) cos (0.25,0)
                                       sin (0.3125,0.5) cos (0.375,0) sin (0.4375,-0.5) cos (0.5,0)
                                       sin (0.5625,0.5) cos (0.625,0) sin (0.6875,-0.5) cos (0.75,0)
                                       sin (0.8125,0.5) cos (0.875,0) sin (0.9375,-0.5) cos (1,0)
                                       sin (1.0625,0.5) cos (1.125,0) sin (1.1875,-0.5) cos (1.25,0)
                                       sin (1.3125,0.5) cos (1.375,0) sin (1.4375,-0.5) cos (1.5,0) -- (1.5,0);
                    \draw[->,thick, blue](1.485,0) -- (1.8,0);
                    \draw[thick, blue](-0.305,0) -- (0.01,0);
                    \node[blue] at (0.72,0.7) {$\omega_r$};
                \end{scope}

                \begin{scope}[xshift=0cm, yshift=4.8cm]
                    \draw[thick, red] (0,0) sin (0.0625,0.5) cos (0.125,0) sin (0.1875,-0.5) cos (0.25,0)
                                       sin (0.3125,0.5) cos (0.375,0) sin (0.4375,-0.5) cos (0.5,0)
                                       sin (0.5625,0.5) cos (0.625,0) sin (0.6875,-0.5) cos (0.75,0)
                                       sin (0.8125,0.5) cos (0.875,0) sin (0.9375,-0.5) cos (1,0)
                                       sin (1.0625,0.5) cos (1.125,0) sin (1.1875,-0.5) cos (1.25,0)
                                       sin (1.3125,0.5) cos (1.375,0) sin (1.4375,-0.5) cos (1.5,0) -- (1.5,0);
                    \draw[->,thick, red](1.485,0) -- (1.8,0);
                    \draw[thick, red](-0.305,0) -- (0.01,0);
                    \node[red] at (0.8,0.7){$\omega_r - \omega_{he} \;\;$};
                \end{scope}

                \begin{scope}[xshift=-1.1cm, yshift=1.9cm]
                    \draw[thick, orange] (0,0) sin (0.0625,0.5) cos (0.125,0) sin (0.1875,-0.5) cos (0.25,0)
                                       sin (0.3125,0.5) cos (0.375,0) sin (0.4375,-0.5) cos (0.5,0)
                                       sin (0.5625,0.5) cos (0.625,0) sin (0.6875,-0.5) cos (0.75,0)
                                       sin (0.8125,0.5) cos (0.875,0) sin (0.9375,-0.5) cos (1,0)
                                       sin (1.0625,0.5) cos (1.125,0) sin (1.1875,-0.5) cos (1.25,0)
                                       sin (1.3125,0.5) cos (1.375,0) sin (1.4375,-0.5) cos (1.5,0) -- (1.5,0);
                    \draw[->,thick, orange](1.485,0) -- (1.8,0);
                    \draw[thick, orange](-0.305,0) -- (0.01,0);
                    \node[orange] at (0.85,0.7){$\omega_{hg}$ \;\;};
                \end{scope}
            \end{tikzpicture}
        };
        \node at (9.55,3) {(c)}; 
    \end{tikzpicture}

    \caption{\justifying (a) Single shot histograms showing the four blobs corresponding to the four computational states. (b) Single shot histograms measured after initialization. (c) Initialization scheme using an intermediate state. We apply two tone pumping along with the fast decay rate of cavity photons to initialize qubits to excited states. }
    \label{fig:ReadoutInitialization}
\end{figure*}

We perform dispersive readout using a 3D copper cavity resonator with frequency of 7.475 GHz and linewidth of 9 MHz. We carry out a single-shot joint readout of the two qubit states using a Josephson traveling parametric-wave amplifier (JTWPA) provided by Lincoln labs \cite{Macklin2015JTWPA} to preamplify the readout signal. The dispersive shifts of qubit A and B are 3 and 6.5 MHz, respectively. By fitting the single-shot histograms as illustrated in Fig. \ref{fig:ReadoutInitialization}(a) with four Gaussian distributions, we extract the population of the four computational states for each experiment. 

Our initialization protocol involves simultaneously driving on the $|g 0\rangle \rightarrow|h 0\rangle$ and $|h 0\rangle \rightarrow|e 1\rangle$ transitions for 25 $\mu \mathrm{s}$, following \cite{Zhang2021fastflux}. Compared with the physical temperature, the high cavity frequency ($7.475 \mathrm{GHz}$) along with a low quality factor of 800 leads to rapid photon loss from the $|e 1\rangle$ state to the $|e 0\rangle$ state as illustrated in \ref{fig:ReadoutInitialization}(c). This prepares qubits A and B in a mixed state with measured excited state populations of 88\% and 87\%. The corresponding single-shot histograms are shown in \ref{fig:ReadoutInitialization}(b). This state initialization enables accurate characterization of gate errors, and we did not focus on improving initialization in this paper.

\section{\label{app:Happ} Model and System Hamiltonian}

In Sec.~\ref{sec:level1}, we presented the truncated system Hamiltonian in the computational subspace. Here, for completeness, we introduce the full Hamiltonian model, which is used in numerical simulations.  We define a two-fluxonium-qubit system with capacitive and inductive couplings as
\begin{equation}\label{eq:full_ham}
\begin{aligned}
\hat{H}_{\mathrm{stat}}&=\hat{H}_{A}+\hat{H}_{B}+\hat{H}_{r}+\hat{H}_{p}\\&+J_{\varphi}\hat{\varphi}_{1}\hat{\varphi}_{2}+J_{C}\hat{n}_{1}\hat{n}_{2}\\&-i[g_{A,r}\hat{n}_{A}(\hat{a}_{r}-\hat{a}_{r}^{\dagger})+g_{B,r}\hat{n}_{B}(\hat{a}_{r}-\hat{a}_{r}^{\dagger})]\\&-i[g_{A,p}\hat{n}_{A}(\hat{a}_{p}-\hat{a}_{p}^{\dagger})+g_{B,p}\hat{n}_{B}(\hat{a}_{p}-\hat{a}_{p}^{\dagger})]\, .
\end{aligned}
\end{equation}
Here
\begin{equation}
\hat{H}_{\alpha}=4E_{C,\alpha}\hat{n}_{\alpha}^2+\dfrac{1}{2}E_{L,\alpha}\hat{\varphi}_{\alpha}^2-E_{J,\alpha}\cos(\hat{\varphi}_{\alpha}-\phi_{\alpha})
\label{eq:qubits}
\end{equation}
describes individual fluxonium qubit $(\alpha=A,B)$, and 
\begin{equation}
    \hat{H}_{i}=\hbar\omega_i\hat{a}^{\dagger}_i\hat{a}_i
\end{equation}
describes the read-out and spurious LC bosonic modes ($i=r,p$). The qubit parameters are listed in Table~\ref{table:para_system}, and the bosonic mode parameters are listed in Table~\ref{table:para_boson}. Here we use the notation $f^{\alpha}_{mn}=\omega^{\alpha}_{mn}/2\pi$ to denote the transition frequency of qubit $\alpha$ between the $m$-th level and $n$-th level. This qubit system is designed so that the static $ZZ$ coupling is suppressed to $\lesssim\SI{3}{\kilo\hertz}$, and that the matrix elements of charge operator $\hat{n}_{\alpha}$ are properly tuned for a $ZX$ (or $XZ$) interaction.

We calculate eigenenergies and eigenstates of the time-independent Hamiltonian~\eqref{eq:full_ham}.  We denote the eigenstates as
$\ket{q_A,q_B,n_r,n_p}$, where $q_A$, $q_B$, $n_r$, and $n_p$ stand for the excitation index of qubits A and B, the readout resonator and the LC modes, respectively.   We assume that the readout and LC modes have no excitations, $n_r=n_p=0$, during the gate execution and use the abbreviated  notations $\ket{q_A,q_B,0,0} = \ket{q_A,q_B}$.

The microwave drive is applied to both qubits and is described by the following term
\begin{equation}\label{eq:DriveHamiltonian}
    \frac{\hat{H}_{\mathrm{mw}}}{h}= 
    \frac{\Bar{\beta}(t)}{2}    \left( e^{-i2\pi f_d t}( \epsilon_A \hat{n}_{A} + \epsilon_B \hat{n}_{B} ) +\text{h.c.}\right),   
\end{equation}
where $\Bar{\beta}(t)$ is a smooth function describing the drive amplitude at frequency $f_d$.  The complex coefficients $\epsilon_{A,B}$ describe the amplitude and phase shift of the drive.
The matrix elements $\bra{q_a,q_b} \hat n_\alpha \ket{q_a',q_b'}$ of the charge operators $\hat n_A$ and $\hat n_B$ are computed in the eigenstate basis of the static Hamiltonian~\eqref{eq:full_ham}.  These matrix elements define the relations between the microwave amplitudes applied to the fluxoniums and the coefficients $\xi_{A,B}^{\pm}$ in Eq.~\eqref{eq:xi_A_pm}:
\begin{equation}
\begin{split}
M_{AA}^\pm  &= -i[\bra{00} \hat n_{A}\ket{10} \pm \bra{01} \hat  n_{A}\ket{11} ],\\
M_{AB}^\pm & = -i [\bra{00}\hat  n_{B}\ket{10} \pm \bra{01}\hat n_{B}\ket{11} ] ,\\
M_{BA}^\pm & = -i [\bra{00}\hat  n_{A}\ket{01} \pm \bra{10}\hat n_{A}\ket{11} ] ,\\
M_{BB}^\pm & = -i [\bra{00}\hat n_{B}\ket{01} \pm \bra{10} \hat n_{B}\ket{11}].
\end{split}
\end{equation}
The values of these coefficients are presented in Table~\ref{tab:MatrixElements}.

\begin{table}[h]
    \centering
\begin{tabular}{cccccccc}
    \hline\hline
     \multirow{2}{*}{Qubit} & $E_{C, \alpha}/h$ & $E_{L, \alpha}/h$ & $E_{J, \alpha}/h$ & $f^{\alpha}_{01}$ & $J_C/h$ & $J_{\varphi}/h$ \\
     & (GHz) & (GHz) & (GHz) & (GHz) & (GHz) & (GHz)\\
    \hline
    $A$ & 0.980 & 0.763 & 5.591 & 0.1472 & \multirow{2}{*}{-0.038} & \multirow{2}{*}{0.0041}\\
    $B$ & 0.993 & 1.155 & 6.271 & 0.2274 & \\
    \hline\hline
\end{tabular}
\caption{\justifying Qubit parameters used in numerical simulations. }
\label{table:para_system}
\end{table}
\begin{table}[t]
    \centering
\begin{tabular}{cccc}
    \hline\hline
     \multirow{2}{*}{Bosonic Mode} & $\omega_{i}/h$ & $g_{A, i}/h$ & $g_{B, i}/h$  \\
     & (GHz) & (GHz) & (GHz) \\
    \hline
    $r$ & 7.4750 & -0.115 & 0.115\\
    $p$ & 3.2165 & -0.182 & 0.208 \\
    \hline\hline
\end{tabular}
\caption{\justifying Bosonic mode parameters used in numerical simulations.}
\label{table:para_boson}
\end{table}

\section{$CX_{\pi}$ gate calibration procedure} \label{sec:CXcal}

\begin{figure*}
    \centering 
    \includegraphics[width=\textwidth]{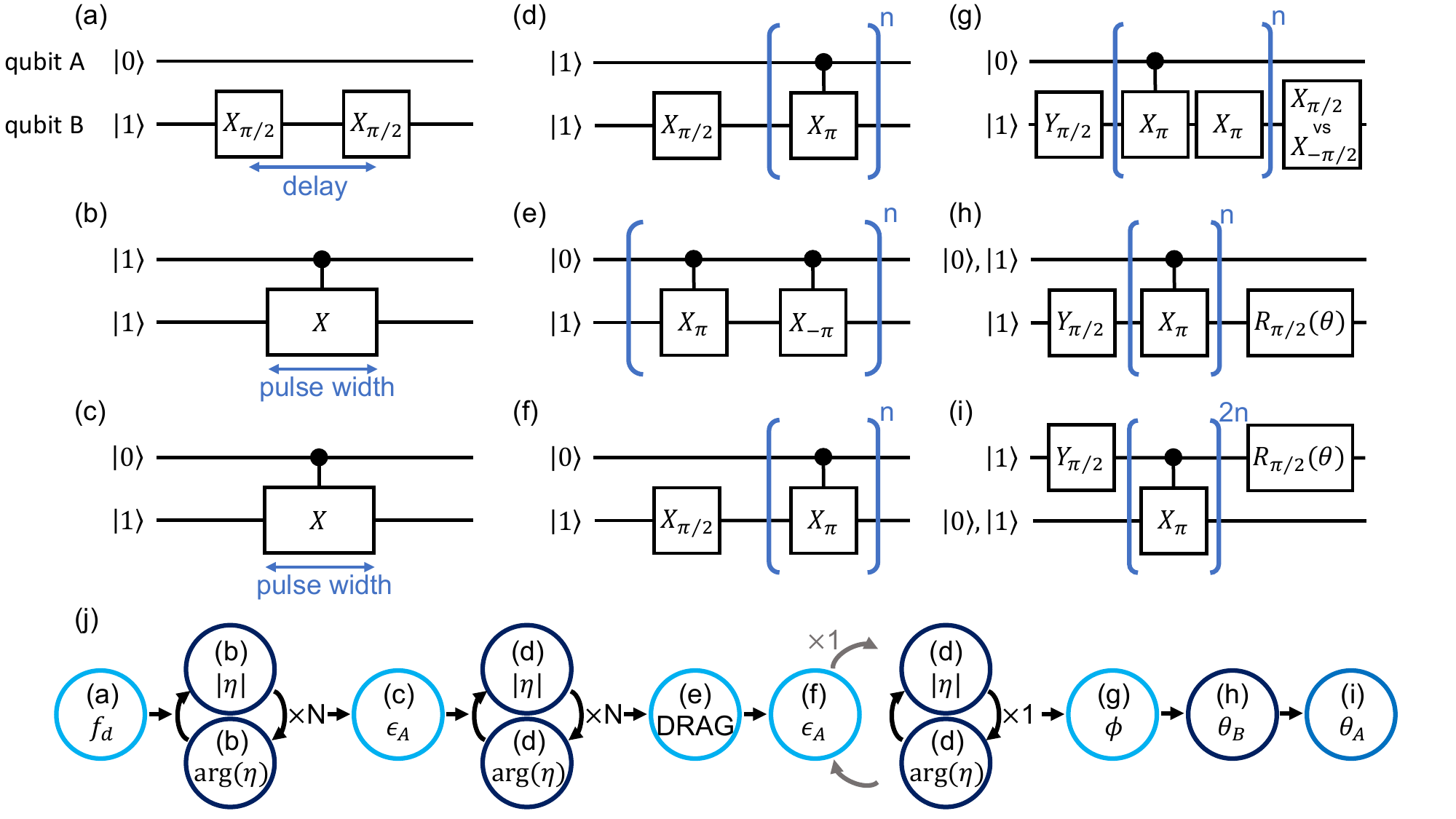}
    \caption{\label{fig:CXflowchart} \justifying Pulse sequences and flow chart for $CX_\pi$ gate tune-up  with corresponding data shown in Fig. \ref{fig:CXflowchartData}: (a) Determining drive frequency by Ramsey experiments. (b)-(c) Rough tuning getting the sub-precise values of the overall amplitude, relative amplitude, and relative phase of the two drives. (b) Calibrating relative amplitude and phase by nulling Rabi frequency to achieve darkening condition. (c) Setting target pulse width and calibrating the corresponding overall amplitude by measuring Rabi oscillation experiments. (d)-(i) Fine tuning with a large varying number of CR pulses to amplify the target error for precise calibration. (d) Calibrating relative amplitudes and relative phase by checking darkening condition, making sure qubit B does not rotate. (e) DRAG calibration by repeating CR pulses with alternating amplitude, making sure that qubit B is driven along the big circle on Bloch sphere. (f) Calibrating the overall drive amplitude by making sure there is no over or under rotations. (g) Calibrating the rotation axis misalignment of CR pulses with respect to the single qubit gate of B, making sure the rotation axis aligns with the X-axis of qubit B. (h),(i) Measuring the single qubit phases of qubit A,B, respectively, needed for compensating phase accumulation due to each CR pulse. Same sequences confirm all accumulated phases are corrected after calibration. (j) Illustration of the full two-qubit gate calibration process. The pulse sequences and corresponding target parameters are provided in the graphical circles. Dark (light) blue circles indicate the dark (bright) transition, while the last normal blue one denotes the control qubit transitions. The iteration number $N$ typically equals to 2 or 3.}
\end{figure*}

\begin{figure*}
    \centering 
    \includegraphics[width=\textwidth]{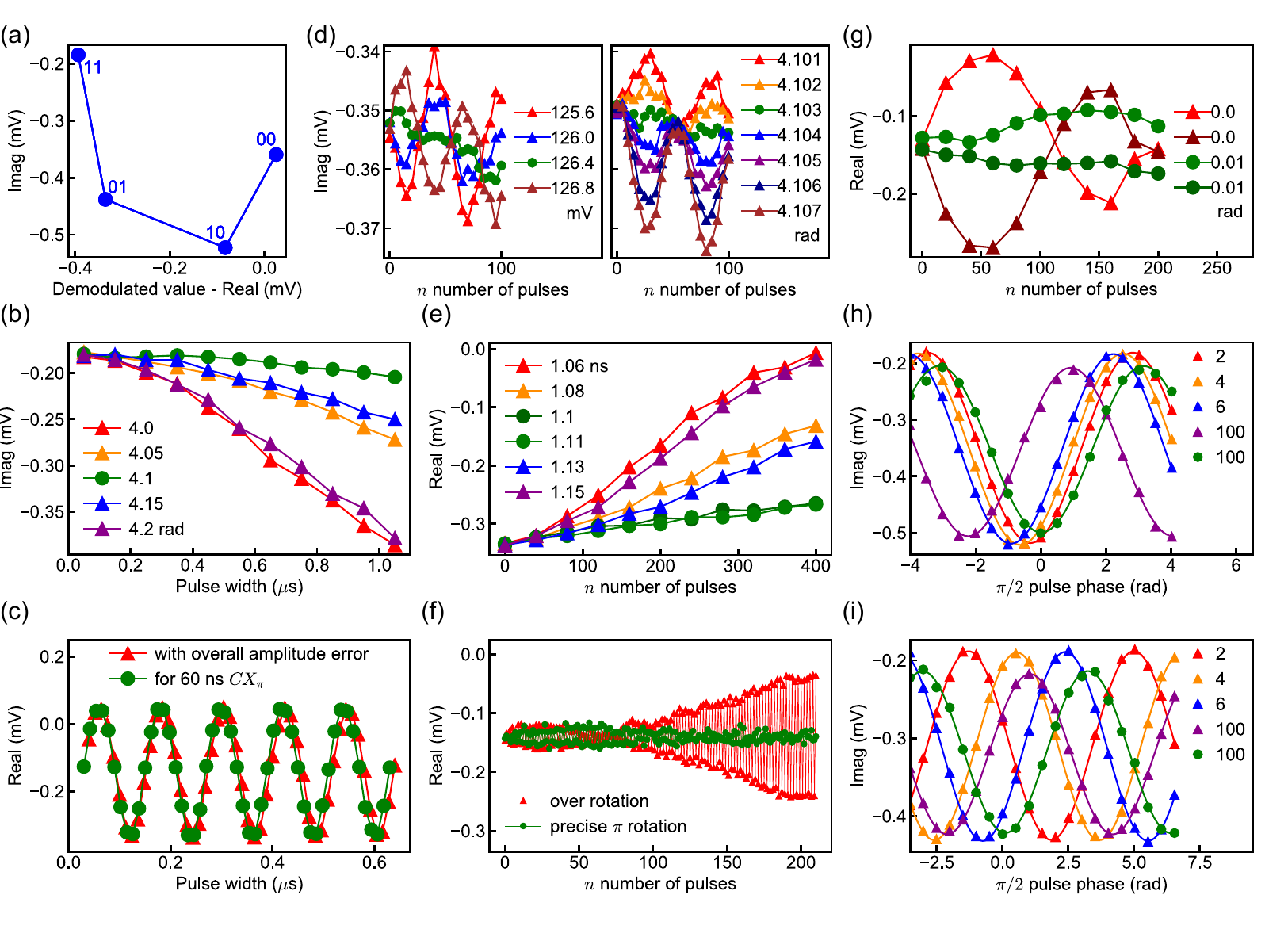}
    \caption{\label{fig:CXflowchartData} \justifying Corresponding $CX_\pi$ gate tune-up data for Fig. \ref{fig:CXflowchart} (b)-(i). The optimized trace (green circle) for each target parameters are shown along with the corresponding error syndrome (red triangle). (a) The demodulated IQ values of the four computational states we prepared, serving as a reference for (b)-(i). We later show the real part for the bright transition $|00\rangle-|01\rangle$ and the imaginary part for the dark transition $|10\rangle-|11\rangle$ in order to show a better contrast. (b) Nulling oscillation frequency to achieve darkening condition, showing the relative phase sweeping as an example, while the relative amplitude sweeping has the same syndrome. (c) Rabi oscillations of a well tuned overall amplitude corresponding to a 60 ns target pulse width and the one with under rotation. (d) Two example data of the iteration through the relative amplitude (left panel) and phase (right panel) sweeping to approach the darkening condition, making sure no detuned Rabi oscillation is observed. (e) Sweeping DRAG and Nulling oscillation frequency to move the rotation axis onto the equator, making qubit B driven along the big circle on Bloch sphere. (f) Repetition of $\pi$ rotations in the bright transition, where precise $\pi$ rotations freeze qubit B population at the fifty-fifty state, while over and under rotations move population up and down with respect to the fifty-fifty state. (g) Comparison of the cases with and without a pulse phase offset between single qubit $\pi$ gates and CR pulses in order to compensate the latitude axis misalignment. The light (dark) color represents the case with a $X_{-\pi/2}$ ($X_{\pi/2}$) at the end. When the rotation axes are well aligned, the oscillations slow down without overlapping. (h),(i) Traces with different number of CR pulse repetition to extract single-qubit phase accumulation for qubit A, qubit B, respectively. When the phases are well compensated, the cosine functions move back to the case of zero CR pulse repetition which has no offset. }
\end{figure*}

To better elaborate our tune-up process, we first define a new variable, $\eta$, with $|\eta| = \epsilon_B/\epsilon_A$ ($\epsilon_A$, $\epsilon_B$ are real), accounting for the relative amplitude of the two local drives. Our AWG needs to apply a phase offset arg($\eta$) between the two drives to address any phase delay due to cables or electronics, making the pulses reaching our sample in-phase with a phase difference of 0 or $\pi$.

In addition, our CR pulse induce phase accumulation due to ac Stark shifts or higher-order CR effect \cite{nesterov2022cnot}, especially after increasing the drive amplitude for faster gates. Accordingly, we further define $\theta_A$ and $\theta_B$ as the phase accumulation of qubit A and B and introduce the evolution matrix of the CR pulse for a $CX_{\pi}$ gate

\begin{equation}
    \begin{pmatrix}
        1 & 0 & 0 & 0\\
        0 & e^{i\theta_B} & 0 & 0 \\
        0 & 0 & 0 & e^{i\theta_A} \\
        0 & 0 & e^{i\theta_A} & 0
    \end{pmatrix}
    .
\end{equation}
Here, we assume that the pulse is well tuned to induce a X rotation, and thus one single variable $\theta_A$ can cover the lower two matrix elements. To compensate $\theta_A$ and $\theta_B$, we apply single-qubit virtual Z gates before ($-\theta_B/2$ on B) and after ($-\theta_B/2$ on B, $\theta_B/2-\theta_A$ on A) the CR pulse to realize a CNOT gate. \cite{nesterov2022cnot,dogan2023two} Note that a CNOT gate is equivalent to a $CX_\pi$ followed by a S gate on qubit A, which can be implemented by a virtual Z gate at no additional cost. In this letter, we benchmark the $CX_\pi$ gate

\begin{equation}
\label{eq:Uid}
    \begin{pmatrix}
        1 & 0 & 0 & 0\\
        0 & 1 & 0 & 0 \\
        0 & 0 & 0 & -i \\
        0 & 0 & -i & 0
    \end{pmatrix}
\end{equation}
instead of a textbook CNOT to point out the dynamics of the conditional rotation governed by Hamiltonian, while the results also apply to a CNOT gate.

Next, we demonstrate the tune-up process determining the eight target parameters, $f_d$, $|\eta|$, arg($\eta$), $\epsilon_{A}$, DRAG coefficient, $\phi$, $\theta_A$, and $\theta_B$, in details. Note that we fix $\eta$ while tuning $\epsilon_{A}$, making it actually an overall amplitude tuning. We should point out that we do not perform any black-box optimizations such as Nelder-Mead or machine learning algorithms. We rely on the following calibration procedure targeting the eight parameters individually, allowing a shorter time cost and higher transparency. Among these tune-up criteria, we put emphasis on achieving the darkening condition with more iterations because the corresponding gate error is larger than the bright transition according to our analysis in Section \ref{Sec:error budget}. Additionally, the limited error amplification of the dark transition discussed later requires longer averaging time to get the precise values of the two corresponding target parameters. Each corresponding tune-up measurement takes more than ten minutes compared with the other steps taking only a few minutes. The whole tuning process takes roughly three hours. Note that our duty cycle for each measurement is over 100 $\mu$s due to hardware limitations, which is longer than the cycle length needed for a few hundreds of 60 ns pulses. The total tuning time can be largely suppressed with shorter duty cycles and automation codes. Fig. \ref{fig:CXflowchartData}(a) shows the demodulated IQ values of the four computational states we prepare, serving as a scale for the latter calibrations.

First, $f_d$ is simply the frequency of the bright transition $|00\rangle-|01\rangle$, which is extracted from Ramsey experiments as shown in Fig. \ref{fig:CXflowchart}(a). We fix this value for all the tuning procedures later.   

Before fine-tuning, we first aim at the sub-precise values of $|\eta|$ and arg($\eta$) to achieve the darkening condition. Since we can only achieve selective darkening with these two parameters being accurate, we iterate through the calibrations sweeping these two values to approach the darkening condition. The measurement we do for this calibration is first setting a rather low $\epsilon_{A}$ to avoid any crazy dynamics under high power and then sweeping the pulse width of our CR drive as shown in Fig. \ref{fig:CXflowchart}(b) to check the rabi oscillation of the dark transition $|10\rangle-|11\rangle$ and minimize the rabi frequeny until there's no oscillation showing a simple decay trace. The initial value of $|\eta|$ can be roughly calculated using Table. \ref{tab:MatrixElements}, so we first fix this rough $|\eta|$ and sweep arg($\eta$) to get the rough arg($\eta$) value. Fig. \ref{fig:CXflowchartData}(b) shows the corresponding data for example. Then we go back to sweep $|\eta|$ for a more accurate value. We repeat this cycle of $|\eta|$ and arg($\eta$) sweeping for a few times and determine the sub-precise values of these two parameters for the next step. 

Then we move on to get the sub-precise $\epsilon_{A}$ for the target pulse width we set. We perform Rabi oscillation measurements as shown in Fig. \ref{fig:CXflowchart}(c) and sweep $\epsilon_{A}$ until our target pulse width induces a $\pi$ rotation for the bright transition. Taking a 60 ns $CX_{\pi}$ for example, one can set a pulse width of 600 ns (multiples of 60 ns) and find the corresponding $\epsilon_{A}$ giving the peak value while sweeping $\epsilon_{A}$. Fig. \ref{fig:CXflowchartData}(c) shows the calibration results and the case with the $\epsilon_{A}$ error.

Starting from this paragraph, we conduct fine-tuning experiments for target parameters using a large varying number of CR pulses $n$ to amplify the target error for precise calibration. For each target parameter, we start with a small $n$, compensate the corresponding error, increase $n$, compensate the error again, and repeat this cycle until $n$ reaches a few hundreds. 

We start with arg($\eta$) and $|\eta|$ for achieving darkening condition using the sequence shown in Fig. \ref{fig:CXflowchart}(d). The $X_{\pi/2}$ pulse at the beginning moves qubit B to superposition state to increase the signal sensitivity to any unwanted rotation with respect to X axis. We iterate through the sweeping of $|\eta|$ and arg($\eta$) to get the precise values after confirming no unwanted rotation is observed. Fig. \ref{fig:CXflowchartData}(d) shows the data of $|\eta|$ and arg($\eta$) sweeping. The two kinds of oscillations observed while sweeping $|\eta|$ and arg($\eta$) can be understood as detuned X and Y rotations, respectively, where the detuning comes from the Stark shift or other second order effects on the dark transition. This detuning is much larger than rabi frequency, making the rotation rate approximately the same as detuning. Thus the rotation angle after one pulse is the same as the phase accumulated, agreeing well with the measured value of $\theta_B$. Note that the error amplification of the unwanted dark transition rotation angle is limited by this small detuning. That is, further increasing the number of repetition pulse can not further amplify the error and give higher accuracy of the target parameters. Accordingly, we simply increase the average number of each measurement, making this calibration the most time consuming step compared with the other target parameters. We do supplementary check to make sure there's also no unwanted rotation with respect to Y axis by changing the $X_{\pi/2}$ at the beginning to $Y_{\pi/2}$. 

Alternate pulse train experiments as shown in Fig. \ref{fig:CXflowchart}(e) amplify the error coming from drive frequency detuning. The trajectory leaves the big circle on the Bloch sphere due to the tilted rotation axis and gets further away if alternating the pulse phases with $\pi$ difference. Thus, parameters such as $f_d$ and DRAG coefficient which is also known as dynamical frequency tuning \cite{Gambetta2011DRAG,Motzoi2009DRAG} can be optimized using this experiments. Here we only sweep and optimize DRAG coefficient for simplicity as shown in Fig. \ref{fig:CXflowchartData}(e). Slightly detuned drive frequency $f_d$ with another well optimized DRAG coefficient can be a new pair giving you a big circle trajectory and even a higher gate fidelity. In this paper, we do not find another error syndrome to optimize $f_d$. Changing $f_d$ of our CR pulse involves the optimization or correction for single qubit gates, making the tune-up process complicated. Instead, we make $f_d$ on resonant with the bright transition to simplify the tune-up procedures.
 
Next, we sweep $\epsilon_{A}$ to achieve precise $\pi$ rotation for the bright transition using Fig. \ref{fig:CXflowchart}(f). The $X_{\pi/2}$ pulse again aim to increase the sensitivity to errors. Fig. \ref{fig:CXflowchartData}(f) shows the result after calibration and the symptom of over and under rotation. The signals of the even and odd number of pulses jump back and forth with respect to the initial value indicating the rotation angle is off from $\pi$. One can also sweep pulse width instead of $\epsilon_{A}$ for shorter gate times as we did for our 48.6 ns $CX_{\pi}$, especially when the microwave components respond nonlinearly to the setted amplitude values. For example, the saturation of IQ mixer under high power could change $|\eta|$ after tuning $\epsilon_{A}$ and thus break the darkening condition. 

In reality, adjusting one of $\eta$, DRAG coefficient, and $\epsilon_{A}$ may change the optimal values of the others, thus one can iterate through Fig. \ref{fig:CXflowchart}(d),(e),(f) for higher accuracy. In practice, we calibrate these three target parameters seperately for only a few times and find that the previous values almost don't change after tuning the latter ones. However, we still carefully design the fine-tuning order of these three parameters to minimize the gate error as discussed below. In theory, the darkening condition can be achieved with accurate $\eta$ and is independent of DRAG coefficient and $\epsilon_{A}$, so we choose $\eta$ as the first fine-tuned parameter. Next, adjusting one of DRAG and $\epsilon_{A}$ slightly change the other one corresponding to a $\pi$ rotation. We select DRAG to tune first because the gate error due to non-optimal DRAG is very small compared with the other two parameters, coming from our calculation based on measured data. For example, a 93 ns $CX_{\pi}$ pulse without DRAG has 50 kHz drive frequency detuning from the frequency corresponding to a rotation axis on the equator, resulting in gate errors on the order of $10^{-5}$ from the tilted rotation axis. We also find the optimal DRAG coefficient has magnitude of 1 ns, and the change of this DRAG after tuning other parameters is less than 0.1 ns. Even though we do not iterate back to tune DRAG again, this tiny change only induce additional error much lower than $10^{-5}$. We thus only calibrate DRAG once and then iterate through $\epsilon_{A}$ and $\eta$ calibration once as shown in Fig. \ref{fig:CXflowchart}(j), where we slightly tune $\eta$ and $\epsilon_{A}$ if necessary. This routine is accurate enough and much faster than simultaneously performing a three-dimensional calibration.

We next tune $\phi$, the last pulse parameter of our CR pulse, giving a pulse phase offset compared with the single-qubit gates of qubit B. Fig. \ref{fig:CXflowchart}(g) considers the rotation axis misalignment between single-qubit gates and our CR pulse. This pulse sequence is sensitive to latitude misalignment, for which we can tune the CR pulse phase to compensate. The corresponding syndrome for latitude misalignment is two overlapped and swapping oscillations shown in Fig. \ref{fig:CXflowchartData}(g), which are for the two cases with $X_{\pi/2}$ or $-X_{\pi/2}$ at the end of the pulse sequence. Note that the signal due to longitude misalignment error is removed by the $X_{\pi/2}$ at the end but can still show up if there's over or under rotation angle off from $\pi$. The corresponding syndrome when there's no latitude misalignment is these two oscillations becoming seperate without overlapping each other. We often see this feature of longitude misalignment along with over/under rotation error due to either CR pulses or single qubit $\pi$ gates while tuning up $\phi$ despite the fact that our CR rotation axis is already on the equator. The reason is that single qubit gate rotation axis is usually a little off from the equator which will be discussed later in Section \ref{sec:1QBcal}. The over/under rotation probably comes from the incomplete calibration of the CR pulse or single qubit gate rotation angle. Despite the error coming from single qubit gates, we are still able to calibrate the latitude misalignment by making sure the two oscillation patterns do not overlap.

The final step is calibrating $\theta_A$, $\theta_B$ using Fig. \ref{fig:CXflowchart}(h), (i), respectively, where we sweep the pulse phase $\theta$ of the $\pi/2$ gate at the end. We extract $\theta_A$ and $\theta_B$ by fitting the data in Fig. \ref{fig:CXflowchartData}(h), (i) to a cosine function with phase offset. Note that the accumulated phase on the control qubit have additional $\pi/2$ from the nature of control rotation governed by system Hamiltoniuan. We exclude this contribution and compare the phase accumulation of different $CX_{\pi}$ gate times, which are inversely proportional to the gate times agreeing with the Stark shift and second order effect. We then compensate $\theta_A$ and $\theta_B$ as mentioned earlier and measured the same experiments again to ensure no phase accumulation is observed. Note that the order of tuning Fig. \ref{fig:CXflowchart}(g) and Fig. \ref{fig:CXflowchart}(h),(i) is important, because the extraction and compensation of $\theta_A$ and $\theta_B$ we mentioned assume that the CR pulse is already a well tuned $X$ rotation. Taking the $\phi$ we measured for example, the CR gate gets additional $2 \times 10^{-5}$ error if the order is flipped.

The graphical flow chart in Fig. \ref{fig:CXflowchart}(j) summarizes the full calibration process.

\section{Single-qubit gate tuning and results} \label{sec:1QBcal}

\begin{figure*}
    \centering 
    \includegraphics[width=\textwidth]{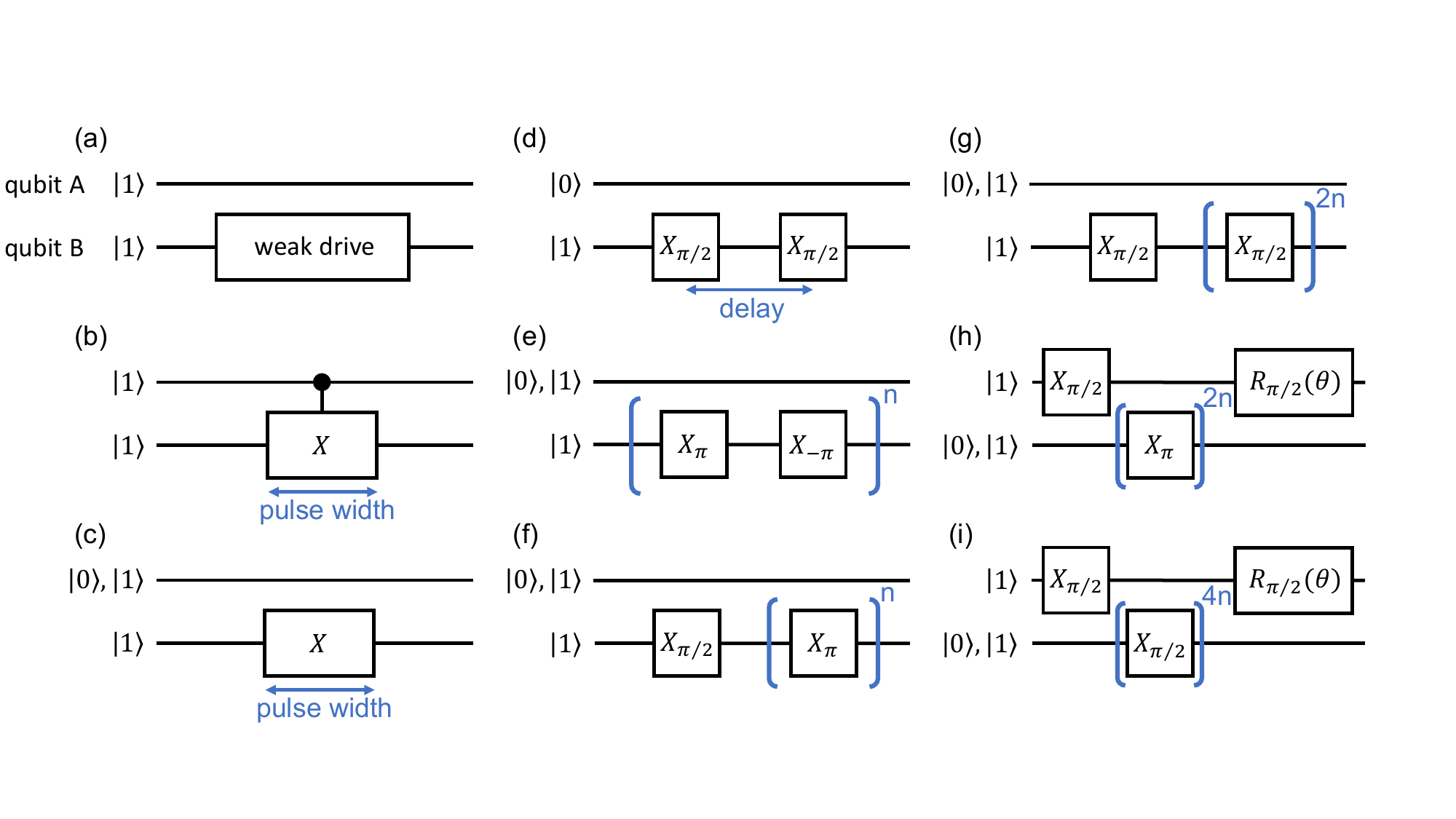}
    \caption{\label{fig:1QBflowchart} \justifying Single-qubit $X_\pi$ and $X_{\pi/2}$ tune up procedures, showing pulse sequences for calibrating qubit B. Calibration for qubit A follows the same sequences with qubit A and B exchanged. (a) Measuring low power spectroscopy of the qubit transitions for drive frequency. (b) Sweeping the relative amplitude and phase of the two drives to achieve darkening condition, determining the relative drive phase needed for the compensation drive on the other qubit. (c) Rabi oscillation experiments. First we calibrate the relative amplitude of the two drives to synchronize the two Rabi rates conditional on the other qubit. Then we sweep the overall amplitude and determine $X_\pi$ pulse width. (d) Ramsey experiments to determine the precise drive frequency. (e)-(i) Fine tuning with a large varying number of $X_\pi$ or $X_{\pi/2}$ pulses to amplify the errors for precise calibration. (e) DRAG calibration by repeating $X_\pi$ pulses with alternating amplitude, making rotation trajectories close to the big circle on Bloch sphere. (f),(g) Calibrating the overall and relative drive amplitudes for $X_\pi$, $X_{\pi/2}$, respectively, making sure there is no over or under rotations. (h),(i) Measuring the accumulated phases coming from the Stark shifts of qubit A due to $X_\pi$, $X_{\pi/2}$ of qubit B, respectively.}
    
\end{figure*}

\subsection{Gate calibration}

Fig. \ref{fig:1QBflowchart} illustrates the calibration of our single-qubit gates with an offset cosine pulse shape without flattop. The target parameters are all the same as the $CX_{\pi}$ case except $\phi$. We first determine the sub-precise $f_d$ by spectroscopy as shown in Fig. \ref{fig:1QBflowchart}(a). We next aim at arg$(\eta)$ because we need simultaneous two port drives to compensate classical crosstalks. Selective darkening requires accurate $\eta$, and we thus determine arg$(\eta)$ by satisfying the darkening condition as shown in Fig. \ref{fig:1QBflowchart}(b). Fig. \ref{fig:1QBflowchart}(c) determines the sub-precise values of $|\eta|$ and $\epsilon_{A}$. Next, we determine the $f_d$ for qubit A and B by Ramsey measurements, making them on resonant with $|00\rangle-|10\rangle$ and $|00\rangle-|01\rangle$ as shown in Fig. \ref{fig:1QBflowchart}(d). We perform the same tune-up sequences as $CX_{\pi}$ to fine tune the DRAG, $|\eta|$, and $\epsilon_{A}$ as illustrated in Fig. \ref{fig:1QBflowchart}(e),(f),(g). We last measured the accumulated phase due to Stark shifts utilizing the sequences in Fig. \ref{fig:1QBflowchart}(h),(i). 

As shown in Fig. \ref{fig:1QBflowchart}, we need to optimize $|\eta|$ and $\epsilon_{A}$ individually for $X_{\pi}$ and $X_{\pi/2}$ instead of setting $\epsilon_{A}$ directly to half because of the saturation of IQ mixer where we observe nonlinear amplitude response while tuning the drive amplitude. Note that applying virtual Z compensation for the phases in Fig. \ref{fig:1QBflowchart}(h),(i) cannot change the results of individual RB but affect the outcome of simultaneous RB. One needs to check single-qubit Clifford decomposition to determine if applying this phase compensation is helpful.

\subsection{Gate characterization}

\begin{figure}
    \includegraphics[width=\columnwidth]{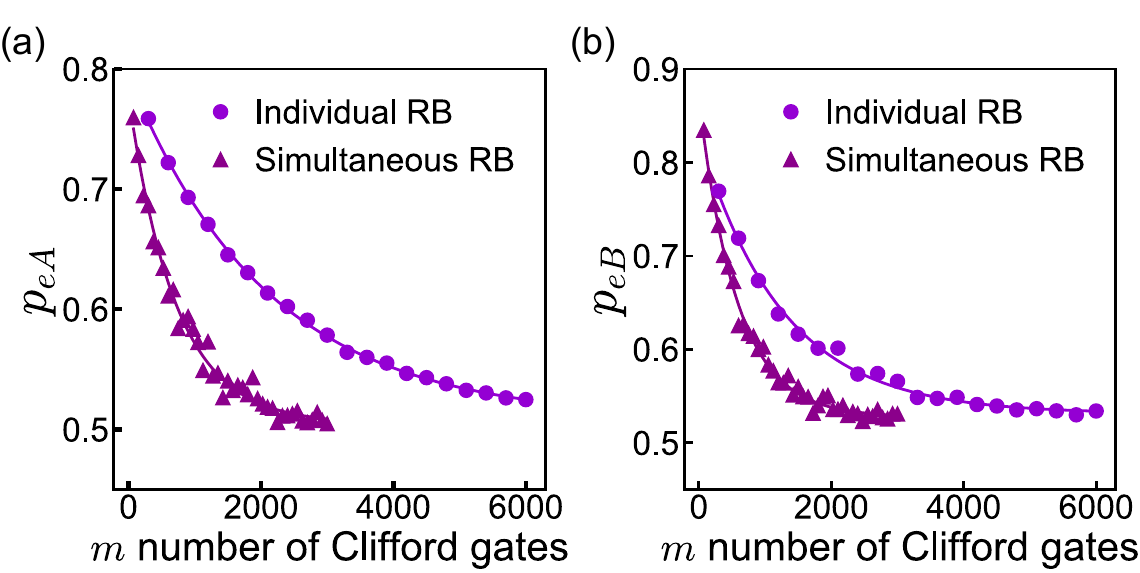}
    \caption{\label{fig:1QB_RB} \justifying Single-qubit gate RB. (a), (b) Individual and simultaneous RB for qubit A, B, respectively.}
\end{figure}

Our decomposition of the single-qubit Clifford group utilizes virtual Z gates, which perform frame updates and change the phase of subsequent X and Y pulses, following the representation described in \cite{xiong2022arbitrary}. Fig. \ref{fig:1QB_RB} shows the results of individual and simultaneous RB for our 61 ns (64 ns) A (B) single-qubit gates. The corresponding fidelities are shown in table \ref{tab:table1}. 

\section{$ZX_{-\pi/2}$ Gate Tuning and results} 

\label{Sec:ZX-90}

\begin{figure*}
    \includegraphics[width=\textwidth]{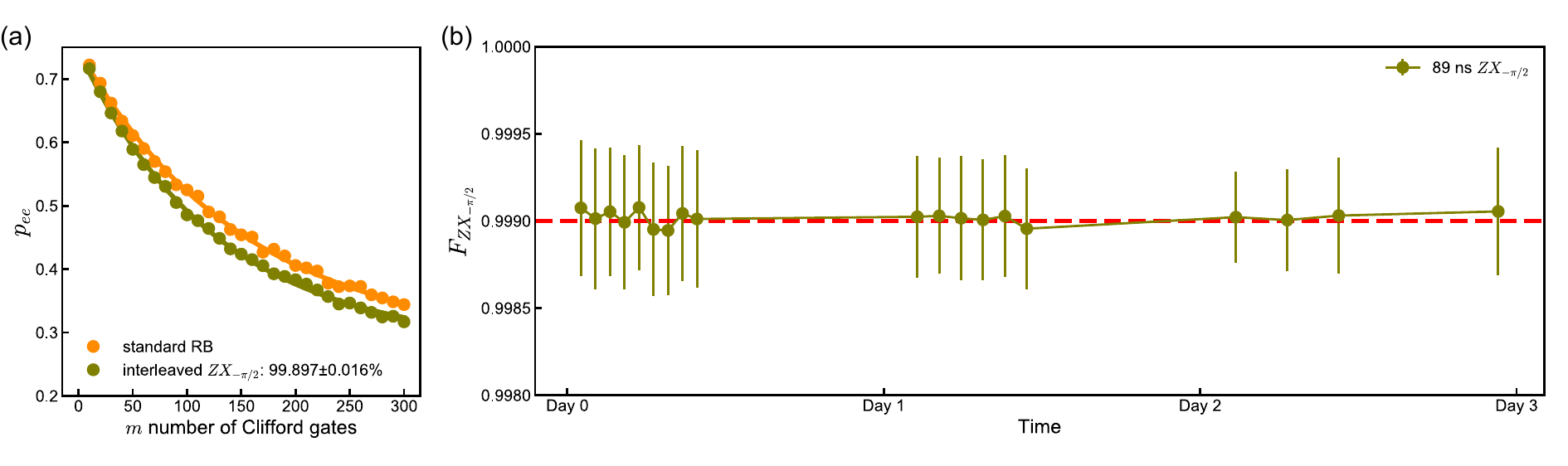}
    \caption{\label{fig:RB_89nsXZ} \justifying $ZX_{-\pi/2}$ IRB results. (a) IRB traces presented with the extracted gate fidelity. (b) Stability demonstration across 3 days.}
\end{figure*}

\subsection{Gate calibration}

For our $ZX_{-\pi/2}$ experiments, qubit A and B are the target and control qubits, respectively, with the CR drive $\hat{n}_{B}$ at the frequency $f_{10}^A$. The calibration follows the tune-up procedures of $CX_{\pi}$ tuning the same eight parameters, the only difference is that the goals of some target parameters change. We again list all the fine-tuned parameters of our $ZX_{-\pi/2}$ pulse and the corresponding goals as below: the relative phase of the two port drives making sure they are in-phase by checking the selectively darkening condition of the $|00\rangle$-$|10\rangle$ transition, the relative amplitude of the two port drives to induce the same rotation speed but opposite rotation direction for the two target qubit transitions, the DRAG coefficient with fixed drive frequency to achieve a big circle trajectory on the Bloch sphere, the overall drive amplitude to ensure a $\pi/2$ rotation angle, the common pulse phase to align the rotation axis to $X$ rotation, and the single-qubit phase accumulation of A and B due to the CR pulse to determine the phase values of virtual Z compensation. Here we only compensate the phase accumulation $\theta_c$ on the control qubit transitions because the two target qubit transitions are under rotation, and thus we observe no phase accumulated. Table. \ref{tab:MatrixElements} indicates the Stark shifts of the two control qubit transitions are the same. Therefore, we found that applying one virtual Z gate with $\theta_c$ phase on qubit B is enough to take care of the accumulated single-qubit phases.

\subsection{Gate characterization}

Fig. \ref{fig:RB_89nsXZ} characterizes the fidelity and stability of our 89 ns $ZX_{-\pi/2}$ . The data points measured on day 2 average over more randomized sequences and thus have longer measurement time and smaller error bars. We preserve the 99.9\% fidelity across 3 days without recalibration interleaved.

\section{Classical crosstalks} \label{Classical crosstalks}

\begin{figure}
    \includegraphics[width=\columnwidth]{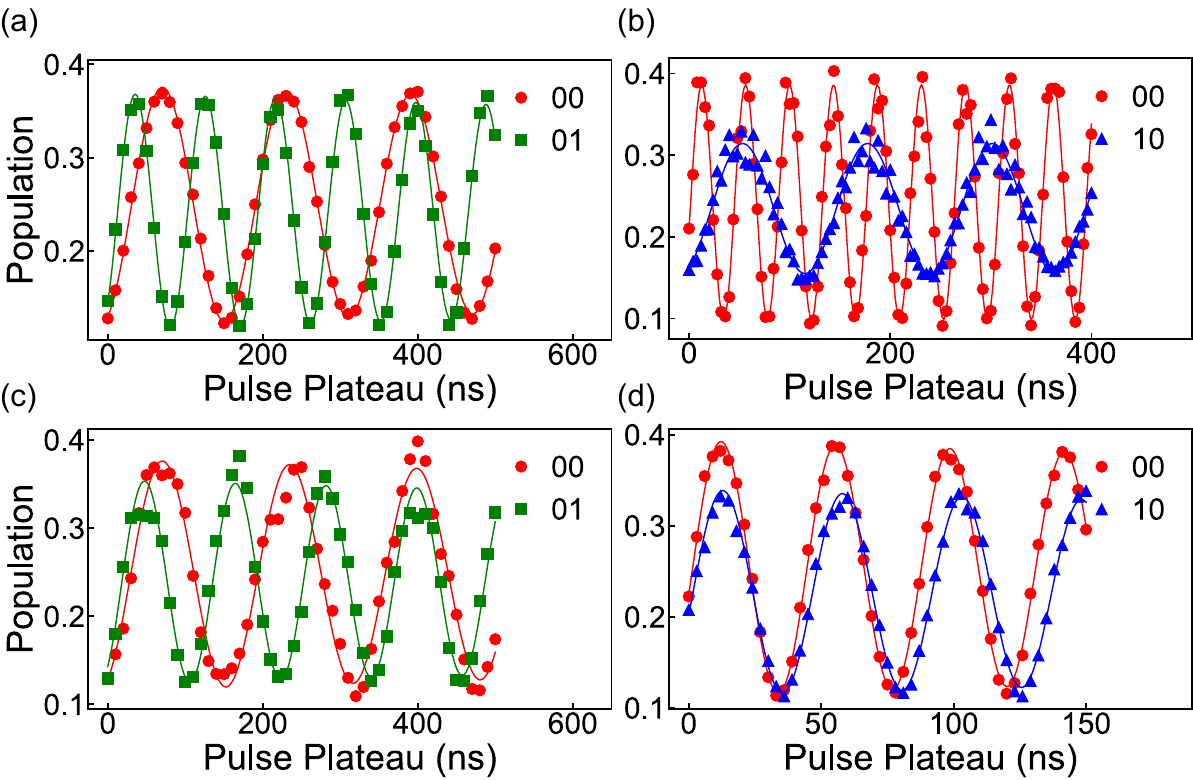}
    \caption{\label{fig:ClassicalCrosstalk} \justifying Rabi oscillation measurements to characterize classical crosstalks. The corresponding port, target qubit frequency, measured Rabi rate, and the calculated $\eta$ are listed below: (a) Port A, $f_{01}^A$: $\Omega_{00-10} = 6.24$ MHz, $\Omega_{01-11} = 11.05$ MHz,  $\eta_A = 0.7$ (b) Port A, $f_{01}^B$: $\Omega_{00-01} = 22.88$ MHz, $\Omega_{10-11} = 8.05$ MHz. $\eta_A = 0.69$ (c) Port B, $f_{01}^A$: $\Omega_{00-10} = 6.14$ MHz, $\Omega_{01-11} = 8.57$ MHz. $\eta_B = 14.99$ (d) Port B, $f_{01}^B$: $\Omega_{00-01} = 23.20$ MHz, $\Omega_{10-11} = 22.20$ MHz. $\eta_B = 14.83$  }
\end{figure}

Note that in a realistic system, each drive from the two ports influences both qubits. Consequently, $|\eta|$ is not merely the ratio of the two drive amplitudes we apply for the two ports. Nonetheless, we can still adjust the ratio of these two amplitudes to construct any combination of $\epsilon_A$ and $\epsilon_B$ we need. To characterize this classical crosstalk effect, we define $\eta_A$ and $\eta_B$ for driving port A and B, respectively. In Fig. \ref{fig:ClassicalCrosstalk}, We drive each port solely and measure Rabi rates to extract $\eta_A$ and $\eta_B$ at the frequency $f_{01}^A$ and $f_{01}^B$. Taking Fig. \ref{fig:ClassicalCrosstalk}(a)
for example, we drive port A at the frequency $f_{01}^A$ and measure the Rabi frequency $\Omega_{00-10}$ and $\Omega_{01-11}$ of the transition $| 00 \rangle-| 10 \rangle$ and $| 01 \rangle-| 11 \rangle$, respectively. Following the expression of Eq. \ref{eq:DriveHamiltonian}, we know

\begin{equation}
    \begin{aligned}
        \frac{\Omega_{00-10}}{\Omega_{01-11}}=\frac{\langle 00|\hat{n}_{A}| 10\rangle+\eta_A\langle 00|\hat{n}_{B}| 10\rangle}{\langle 01|\hat{n}_{A}| 11\rangle+\eta_A\langle 01|\hat{n}_{B}| 11\rangle}.
    \label{eq:IX}
    \end{aligned}
\end{equation}
Incorporating with Table \ref{tab:MatrixElements}, we can calculate the value of $\eta_A$ for the frequency $f_{01}^A$. Similarly, we can also drive port A at the frequency $f_{01}^B$, port B at $f_{01}^A$, and port B at $f_{01}^B$ with corresponding results shown in Fig. \ref{fig:ClassicalCrosstalk}(b),(c), and (d), respectively.

\section{Error Budget for Selective-darkening Gates and Single-qubit X Gates}\label{app:budget}

To evaluate the sources of coherent error for the selective-darkening gate, we run simulations where qubit A is the control and qubit B is 
the target. The drive generates a $\pi$ rotation between states $\ket{00}$ and $\ket{01}$ (the bright transition) and keeps the state in the $\ket{10}\leftrightarrow\ket{11}$ subspace fixed (the dark transition). We compute the evolution operator for the driven system using QuTiP package~\cite{Johansson2012,Johansson2013}. We  project it to the computational subspace $4\times 4$ and obtain matrix $\hat U_{\rm sim}$ to evaluate the gate fidelity for the CX gate given by $\hat U_{\rm ideal}$~\eqref{eq:Uid} using the standard expression for the gate fidelity~\cite{Pedersen2007}
\begin{equation}
    F = \dfrac{\Tr(\hat{U}_{\rm sim}^{\dagger}\hat{U}_{\rm sim})+\abs{\Tr[\hat{U}_{\rm ideal}\hat{U}_{\rm sim}]}^2}{20}\,.
\end{equation} 
The fidelity for the CNOT gate can be expressed in terms of the error and leakage  probabilities as~\cite{nesterov2022cnot} 
\begin{equation}
1-F = \mathcal{E}_{\mathrm{ctrl}}^0 + \mathcal{E}_{\mathrm{ctrl}}^1+ \mathcal{E}_{\mathrm{dark}} +\mathcal{E}_{\mathrm{bright}}+\mathcal{E}_{\mathrm{leak}}\,. 
\end{equation}
Here, we identify five contributions to the gate error:
\begin{itemize}
    \item 
    $\mathcal{E}_{\mathrm{ctrl}}^0= (P_{00\to 10}+P_{10\to 00} + P_{01\to 10}+P_{10\to 01})/5$, the probability of control qubit flip when target qubit is in  state $\ket{0}$ plus the probability of swapping;
    \item $\mathcal{E}_{\mathrm{ctrl}}^1= (P_{01\to 11}+P_{11\to 01} + P_{00\to 11}+P_{11\to 00})/5$, the probability of control qubit flip when target qubit is in state $\ket{1}$ plus the probability of double (de)-excitation;
    \item $\mathcal{E}_{\rm dark} =(P_{10\to 11}+P_{11\to 10})/5$, the probability of the dark transition $\ket{10}\leftrightarrow\ket{11}$;
    \item $\mathcal{E}_{\rm bright} = (P_{00\to 00}+P_{01\to 01})/5 $, the probability of imperfect bright Rabi flip between states 
    $\ket{00}\leftrightarrow\ket{01}$;
    \item $\mathcal{E}_{\mathrm{leak}} = 1-\Tr \{\hat{U}_{\rm sim}^{\dagger} \hat{U}_{\rm sim}\}/4$, the leakage probability.
\end{itemize} 
The error probabilities above are determined by the matrix elements of the simulated evolution operator $\hat{U}_{\rm sim}$:
\begin{equation}
P_{ab\to a'b'} = |\bra{ab}\hat{U}_{\rm sim}\ket{a'b'}|^2
\end{equation}
We group the four error probabilities $P_{00\to 10}$, $P_{10\to 00}$, $P_{01\to 10}$, $P_{10\to 01}$ into $\mathcal{E}_{\rm ctrl}^0$ because we observe that the difference between these probabilities is less than $10\%$ of their values. The $\ket{00}\to\ket{10}$ and $\ket{10}\to\ket{00}$ process are symmetrical and therefore are expected to have very similar probabilities. The $\ket{01}\leftrightarrow\ket{10}$ transition can be broken down into the control qubit flip $\ket{00}\leftrightarrow\ket{10}$ and the bright transition $\ket{01}\leftrightarrow\ket{00}$. Since the bright transition probability is close to 1 with proper drives, the order of magnitude of the probability of swapping transition should thus be determined by the probability of control qubit flip $\ket{00}\leftrightarrow\ket{10}$ and therefore is very close to $P_{00\to 10}$ and $P_{10\to 00}$ numerically. For similar reason, $P_{01\to 11}$, $P_{11\to 01}$,$P_{00\to 11}$, $P_{11\to 00}$ are grouped into $\mathcal{E}_{\rm ctrl}^1$.

\begin{figure}
    \includegraphics[width=0.4\textwidth]{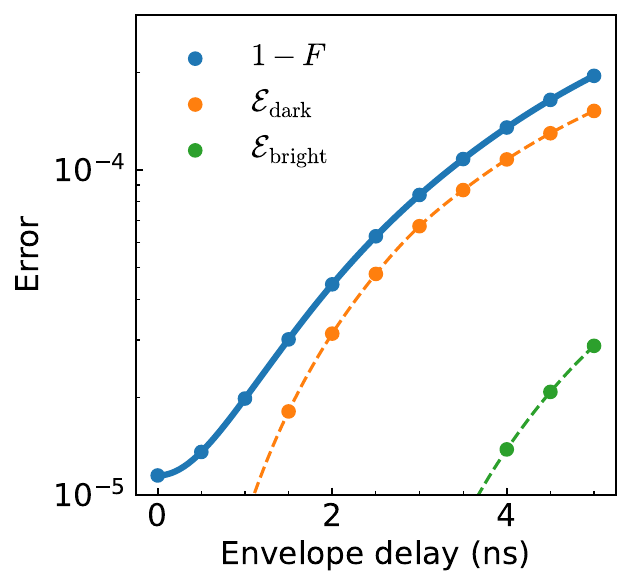}
    \caption{\justifying Error budget of 60 ns $CX_\pi$ gate for different drive envelope delay. The solid line is the coherent error of the optimized gate. The dashed lines are the decomposition of the coherent error into different error channels.}
    \label{fig:envelope_delay}
\end{figure} 

Since our model has two qubits and two resonator modes, and the truncated dimension of each mode should be no less than 5 to have numerically reliable simulation results, it is hard to numerically solve the master equation to simulate decoherence. Therefore, we use the analytical expression to estimate incoherent error:~\cite{tripathi_2019}
\begin{equation} \label{eq:IncoherentError}
    \Delta F_{\rm incoherent}\simeq\dfrac{1}{5}\dfrac{t_{\rm gate}}{T_{1,A}}+\dfrac{1}{5}\dfrac{t_{\rm gate}}{T_{1,B}}+\dfrac{2}{5}\dfrac{t_{\rm gate}}{T_{2,A}^E}+\dfrac{2}{5}\dfrac{t_{\rm gate}}{T_{2,B}^E}.
\end{equation}
Here $t_{\rm gate}$ is the gate time of the CX gate, and $T_1$ and $T_2^E$ are taken from Table~\ref{tab:table1}.

The simulated coherent error and the estimated incoherent error are both presented in Fig.~\ref{fig:cnot_sim}. The curves of the two types of errors cross each other around $\SI{40}{\nano\second}$, and for gate time $\gtrsim\SI{50}{\nano\second}$, the incoherent error becomes more than an order of magnitude larger than the numerically optimized coherent error. For the coherent gate simulation, we vary drive frequency, drive amplitudes, and the ratio between qubit A and qubit B drives to optimize the $CX_\pi$ gate. We use system parameters from Table~\ref{table:para_system} and Table~\ref{table:para_boson} and evaluate the coherent gate fidelity for the gate time ranging from 30 to 100ns. For shorter gate times, $t_{\textrm gate}\leq\SI{40}{\nano\second}$ in simulations, the  errors involving control qubit flip dominate. Therefore, we choose to prioritize minimizing the dark transition error and bright rotational error, which are orders of magnitude lower than the control qubit flip error. To include this criterion in the optimization process, we minimize a combination of the total gate error and, with additional weights, the errors associated with the control qubit transitions, $\mathcal{E}_{\textrm ctrl}^{0,1}$. 
The drive parameters resulting from this optimization do not correspond to the exact maximal gate fidelity but still generate a gate with fidelity of the same order of magnitude. The control qubit flip error also remains of the same order of magnitude for $<10\%$ change in drive parameters. At the same time, by optimizing the dark transition error and rotational error, we can have a better understanding of the optimization limit on these two error channels.  We essentially prioritize conditions for the perfect control of the target gate in terms of $\xi_B^\pm$, without fine-tuning of $\xi_A^\pm$, consistent with the calibration procedure described in Appendix~\ref{sec:CXcal}.

Figure~\ref{fig:cnot_sim} shows that the coherent fidelity of the selective-darkening gate is mostly determined by the flip error on the control qubit. As the gate time gets shorter, the drive amplitudes increase, thus inducing more unwanted rotation on the control qubit and enhancing the coherent error. The dark transition error also increases with drive amplitudes, but overall is an order of magnitude lower than the control qubit flip error and is suppressed by proper choice of the ratio between the drive amplitudes on qubits A and B. 
In our simulations, we are able to reduce the bright error by optimization of the drive amplitudes and frequency to achieve high-fidelity transition between states $\ket{00}$ and $\ket{01}$. The dark error also drops below  $10^{-7}$ for the gate time exceeding $\geq\SI{60}{\nano\second}$.  In this case, the error is dominated by the off-resonant transitions of the control qubit. This error remains dominant at shorter gate times, as the drive amplitude increases and the off-resonant transitions increase even further.  
Finally, for all these gate times, the leakage error is well below $10^{-7}$; therefore, is not present in the figure. The negligible leakage is a result of fluxonium qubits level structure, in which the significant separation between the 0-1 transition frequency and 1-2 transition frequency effectively suppresses leakage out of computational subspace during the drive.

The cable length difference between our two port drives is a few tens of centimeters, inducing the drive envelope delay of a few nanoseconds. We do not calibrate against this delay in our experiments, so we further simulate this effect for our 60 ns $CX_\pi$ as shown in Fig. \ref{fig:envelope_delay}. We sweep the delay up to 5 ns and observe a significant rise of the dark transition error to the order of $10^{-4}$. This agrees with the measured upper bound of $2 \times 10^{-4}$ for coherent errors. The bright transition error can also go above $10^{-5}$, while the control qubit flip and leakage errors are still well below $10^{-5}$ under this effect and are not shown in Fig. \ref{fig:envelope_delay}. 

We next simulate single-qubit X gates to investigate the additional error sources of simultaneous single-qubit gates compared with individual single-qubit gates. The lowest simulated coherent errors of the individual X gate on qubit A, the individual X gate on qubit B, and the simultaneous X gates on both qubits are $8 \times 10^{-5}$, $8 \times 10^{-6}$, and $6 \times 10^{-4}$, respectively. For comparison, we estimate the coherent error upper bounds based on the RB measurements in Appendix \ref{sec:1QBcal} and the estimated incoherent errors following \cite{Abad2022universal}. The coherent error upper bounds are $1 \times 10^{-4}$, $2 \times 10^{-4}$, and $8 \times 10^{-4}$ for the individual gate on qubit A, the individual gate on qubit B, and the simultaneous gate pair, respectively. Note that these bounds are estimated for the average errors of the whole single-qubit Clifford group. The estimated errors agrees well with the bounds, and we suspect that the relatively large difference between the estimated values based on experimental data and the simulated values of the individual X gate on qubit B comes from improper calibration and the swap errors observed while tuning up. Following the experiments, we don't optimize the drive frequencies of individual gates for the simulated values above, while the errors of the individual X gate on qubit A and the simultaneous X gate pair can be lowered down to $1 \times 10^{-5}$ and $4 \times 10^{-4}$, respectively, by doing so. In addition, we only optimize the individual X gates and directly apply both simultaneously to achieve the simultaneous X gate pair above. If we further optimize the two drive frequencies of the X gate pair, the simulated error can be suppressed to $7 \times 10^{-5}$, indicating that there's still room to improve single-qubit gates. The difference between the optimized drive frequencies of the individual X gates and the simultaneous X gate pair is above 100 kHz, agreeing with an error over $10^{-4}$ due to detuning. We interpret this behavior as an indicator that the induced Stark shifts contribute significantly to the coherent error of simultaneous gates. Larger qubit-qubit detuning in future devices can mitigate this effect and the swap error while driving two qubits simultaneously.

\nocite{*}

\bibliography{apssamp}
\end{document}